\documentclass[11pt]{article}
\usepackage{jheppub} % for details on the use of the package, please
\hypersetup{colorlinks=true,citecolor=blue,linkcolor=blue,breaklinks=true}
\usepackage{epsfig, amssymb}
\usepackage{graphicx,epsfig}
\usepackage{epstopdf}
\usepackage{graphicx}
\usepackage{caption}
\usepackage{subfig}
\usepackage{color}
\usepackage{amsmath}
\usepackage[normalem]{ulem}
\usepackage{caption}
\usepackage{float}

\hypersetup{
  colorlinks=true,
  linkcolor=red, 
  filecolor=magenta,   
  urlcolor=blue,
  citecolor=blue,
}
\usepackage[T1]{fontenc}

\newcommand{\cmb}[1]{{\color{blue}{#1}}} 
 
\newcommand{\rmd}{\textrm{d}}

\begin{document}

\preprint{CTPU-PTC-26-18}

%\begin{flushright}
%CTPU-PTC-26-18
%\end{flushright}
% ~\vspace{1cm}\\

%%%%%%%%%%%%%%%%%%%%%%%%%%%%%%%%%%%%%%%%%%%%
% Titlepage
%%%%%%%%%%%%%%%%%%%%%%%%%%%%%%%%%%%%%%%%%%%%
 
\title{ Cosmological implications for hairy black holes via spontaneous symmetry breaking:
Are Hairy Black Holes Primordial?}

\author[a,b]{Suruj Jyoti Das}
\author[a,c]{Miok Park}

\affiliation[a]{Particle Theory  and Cosmology Group, Center for Theoretical Physics of the Universe,
Institute for Basic Science (IBS),
 Daejeon, 34126, Korea}
\affiliation[b]{Faculty of Physics, University of Warsaw, Pasteura 5, 02-093 Warsaw, Poland}
\affiliation[c]{Department of Physics Education, Pusan National University, Busan 46241, Republic of Korea}

\emailAdd{Suruj-Jyoti.Das@fuw.edu.pl}
\emailAdd{miokpark76@gmail.com}

\vspace{1cm}

% \begin{center}
% Suruj Jyoti Das, Miok Park

% \vspace{0.5cm}{\small{IBS,\\
% Republic of Korea}}
% \end{center}
%\vspace{1cm}

\abstract{We investigate whether hairy black holes generated through spontaneous symmetry breaking in Einstein–Scalar–Gauss–Bonnet (ESGB) theory, involving a complex scalar field with a global \(U(1)\) symmetry, can be compatible with cosmological evolution. To this end, we introduce the ESGB theory with a scalar self-interaction that becomes relevant on cosmological scales while remaining negligible near the black hole. Owing to the time dependence of the GB term on cosmological scales, the scalar field dynamics in the evolving FLRW background differ qualitatively from those in the nearly static black hole background. In particular, for scalar–GB couplings compatible with hairy black hole formation, the effective potential supports a symmetry-broken vacuum throughout inflation. However, after inflation, a decelerated expansion changes the sign of the GB term, temporarily making the effective potential unbounded from below. As the GB contribution subsequently decreases, the scalar self-interaction eventually dominates and restores the symmetry. Within this schematic framework, we derive stringent constraints on the coupling strengths, the cutoff scale, and the black hole mass, which primarily arise for avoiding efficient tachyonic amplification of the scalar field perturbations during the unbounded phase. For cutoff scales compatible with both cosmological evolution and scalar hair formation, we find that only ultralight black holes with masses of the order of a few grams can develop scalar hair, identifying them as hairy primordial black holes.}

\maketitle

\setcounter{page}{1}

%%%%%%%%%%%%%%%%%%%%%%%%%%%%%%%%%%%%%%%%%%%%
% Table of Contents
%%%%%%%%%%%%%%%%%%%%%%%%%%%%%%%%%%%%%%%%%%%%

%\tableofcontents

%%%%%%%%%%%%%%%%%%%%%%%%%%%%%%%%%%%%%%%%%%%%
% Body of Paper
%%%%%%%%%%%%%%%%%%%%%%%%%%%%%%%%%%%%%%%%%%%%

\newpage

%--------------------- section --------------------------------------------

%--------------------- section --------------------------------------------
\section{Introduction}
%--------------------- section --------------------------------------------

% \begin{itemize}
% 	\item Aim: To check if the Hairy Black Hole (BH) solutions by Spontaneous Symmetry Breaking (SSB) in the presence of Scalar-Gauss-Bonnet Coupling \cite{} are valid in a cosmological FLRW background, and see possible implications and constraints.
% \end{itemize}

Advances in astronomical instrumentation and survey capabilities have opened new observational windows on the Universe and are producing astrophysical data sets of unprecedented scale. Accordingly, recent major astronomical roadmaps have identified time-domain and multi-messenger astronomy as key priorities for the coming decade, with particular emphasis on next-generation gravitational-wave observations and the theoretical and computational capabilities required to interpret them. In light of these developments, GW250114 \cite{LIGOScientific:2025rid} stands out as one of the clearest black hole ringdown signals observed to date, with a reported signal-to-noise ratio of approximately 76–80, depending on the analysis. This event offers an opportunity to extract quasinormal modes and their overtones with high precision from the post-merger ringdown signal. Because the ringdown phase directly probes the strong-field dynamics of the remnant black hole, its analysis can enable stringent tests of general relativity, provide detailed information about the properties of Kerr black holes, and offer observational tests of the no-hair theorem.

One important motivation for testing general relativity is that, although it has been extraordinarily successful, it does not by itself provide a complete explanation for dark matter, dark energy, or the origin of cosmic inflation \cite{Clowe:2006eq,Linde:1983gd,Starobinsky:1980te}. This has motivated the development of numerous alternative and extended theories of gravity \cite{Lovelock:1971yv,Kobayashi:2011nu,Copeland:2012qf,Babichev:2022djd,Dorlis:2023qug}, in which the Einstein–Hilbert action is modified either through additional geometric terms or through nonminimal couplings to new fields. Among these theories, models involving the Gauss–Bonnet (GB) invariant are of particular interest. In four-dimensional spacetime, the GB term is a topological invariant and contributes only a total derivative \cite{tHooft:1974toh}. However, when it is nonminimally coupled to a scalar field, it can affect the dynamics while still yielding equations of motion that contain no derivatives higher than second order \cite{Lovelock:1971yv,Kobayashi:2011nu,Copeland:2012qf}. Moreover, GB corrections arise naturally in the low-energy effective actions of several string-theoretic constructions \cite{Zwiebach:1985uq,Gross:1986mw}.

Einstein–Scalar–Gauss–Bonnet (ESGB) gravity introduces a nonminimal coupling between a scalar field and the GB invariant. This coupling enables the theory to evade some of the assumptions underlying the standard no-hair theorems. The issue was first investigated in \cite{Antoniou:2017acq} and, for a dilatonic coupling, in \cite{Lee:2018zym}, based on Bekenstein’s arguments \cite{Bekenstein:1972ny,Bekenstein:1995un}. It was subsequently revisited in \cite{Papageorgiou:2022umj}, where a subtle aspect of the original treatment was clarified and a more complete proof was presented. The existence of these solutions has motivated further studies of the mechanisms by which scalar hair can develop around initially bald black holes. One prominent mechanism is spontaneous scalarization \cite{Silva:2017uqg}, in which a tachyonic instability, generated by an effectively negative squared mass for scalar perturbations, drives the growth of a nontrivial scalar configuration from an initially constant scalar background. The stability of the resulting scalarized black holes, however, must be assessed separately in each model and parameter regime \cite{Blazquez-Salcedo:2018jnn, Minamitsuji:2018xde, Minamitsuji:2023uyb, Ballesteros:2025wvs}. An alternative mechanism, explored in \cite{Latosh:2023cxm,Hyun:2024sfv}, employs a scalar coupling function containing quadratic and quartic interaction terms and proposes a theoretical mechanism through which a phase transition from bald to hairy black holes may occur via spontaneous symmetry breaking. An important advantage of this mechanism is that the final hairy configuration can be dynamically stable. Moreover, it provides a realization of spontaneous symmetry breaking in an asymptotically flat spacetime. Although spontaneous symmetry breaking is a central organizing principle in fundamental physics and cosmology, it has rarely been investigated as a mechanism for black hole spacetime in asymptotically flatness. The proposed formation of hairy black holes through spontaneous symmetry breaking in ESGB gravity suggests that spontaneous symmetry breaking, a fundamental mechanism in particle physics and cosmology, may also operate in black hole systems in our Universe and provide a viable mechanism for the development of scalar hair.

It would therefore be interesting to investigate whether this mechanism can be consistently embedded within cosmological history. The cosmological consistency of hairy black hole solutions via spontaneous scalarisation has been explored earlier in Refs. \cite{Anderson:2016aoi, Franchini:2019npi, Anson:2019uto, Antoniou:2020nax, Erices:2022bws, Babichev:2024txe}. Particularly, in Refs. \cite{Anderson:2016aoi, Franchini:2019npi, Anson:2019uto} for a quadratic GB coupling function, it has been shown that parameters that lead to scalar hair in high-curvature local environments of astrophysical black holes in the late Universe, also cause an equivalent tachyonic instability in the early Universe.  Without extreme fine-tuning of cosmological initial conditions, these scalarisation models would hence be incompatible with observations and be effectively ruled out. However, it may not be sensible to stretch these theories all the way up to the high energy and small length scales of the early Universe, when a more complete UV theory may be operational. Ref. \cite{Erices:2022bws} addresses this issue by modifying the scalarisation theory from an effective field theory (EFT) perspective, and adding other terms in the theory with $\mathcal{O}(1)$ Wilson coefficients. It was found that for the length scales corresponding to the scalarisation of astrophysical black holes, the EFT of the GB quadratic theory breaks down around the time of Big Bang Nucleosynthesis (BBN). In general, it was concluded that higher the cutoff scale of the GB quadratic theory, the further back in time the EFT is appropriate. However, such higher cutoff scales have not been explored earlier to the best of our knowledge, mostly keeping in mind the observational relevance of the lower mass of the scalarisable objects.

In this paper, we investigate whether the gravitational theory with a complex scalar field proposed in Refs. \cite{Latosh:2023cxm,Hyun:2024sfv}, together with its associated spontaneous symmetry-breaking mechanism for generating hairy black holes, can be consistently embedded within cosmological evolution. Working within an ESGB theory with a global U(1) symmetry, characterized by quadratic and quartic couplings of the scalar field to the GB invariant, we construct an effective description of black holes in the early Universe shortly after inflation. In particular, we identify the conditions that trigger the phase transition from bald to hairy black holes and derive the constraints on the theory required for cosmological consistency. We also introduce a bare quartic self-interaction for the scalar field to stabilize its cosmological evolution after inflation, while assuming that this additional interaction has a negligible effect on the black hole solution. Although our gravity action might not be UV complete, we regard it as an EFT valid below the cutoff scale $\Lambda_{\textrm{UV}}$.

%In this paper, we investigate whether the gravitational theory proposed in Ref. \cite{Latosh:2023cxm, Hyun:2024sfv} with a complex scalar field, and the associated spontaneous symmetry breaking mechanism for hairy black holes can be consistently embedded within cosmological evolution. Followed the setting the ESGB theory with a global U(1) symmetry, characterized by quadratic and quartic scalar couplings to the GB invariant, we study an effective description of black holes in the early Universe shortly after inflation. In particular, we identify the conditions imposed on black holes and derive constraints on the theory that are required for cosmological consistency. We further introduce a bare quartic self-interaction for the scalar field in order to stabilize its cosmological evolution after inflation, while assuming that this additional interaction has a negligible effect on the black hole in the near-horizon region. Although our gravity action might not be UV complete, we regard it as an EFT valid below a cutoff scale $\Lambda_{\textrm{UV}}$.

 On cosmological scales, the GB invariant is determined by the FLRW geometry and evolves dynamically with the cosmological background. Our above setup implies that the scalar field resides in a symmetry-broken vacuum during inflation and remains stabilized there. However, immediately after inflation, the effective scalar potential changes as a consequence of the sign change of the cosmological GB invariant. As a result, the scalar field, which had previously settled in the symmetry-broken vacuum, can be driven towards a local maximum of the post-inflationary effective potential, and subsequently undergo a tachyonic growth, analogous to that found earlier in Refs. \cite{Anderson:2016aoi, Franchini:2019npi, Anson:2019uto}. In order to avoid efficient tachyonic growth which may ruin the cosmological history, we find strong constraints on our cutoff scale $\Lambda$ and the Schwarzschild black hole mass. We essentially require the cutoff scale to be much higher, close to the inflationary Hubble scale, leading to Schwarzschild black hole mass in the ultralight range of a few grams. We thus entertain, perhaps for the first time, the possibility of Hairy Primordial Black Holes (PBHs). Although such ultralight PBHs would likely have evaporated by the present day and thus cannot be directly observed, their existence could nevertheless have had important implications for the early Universe \cite{Hawking:1971ei, Carr:1974nx}. In this work, we explore the intriguing possibility that such ultralight PBHs possessed scalar hair. A detailed investigation of their phenomenological properties, including Hawking radiation, is left for future work.

The paper is organised as follows. In Sec. \ref{sec:setup}, we introduce our framework within ESGB theory, examine the effective scalar potential in black hole and FLRW backgrounds, and identify the conditions required for consistency with both hairy black hole solutions and cosmological evolution. In Sec. \ref{Sec3}, we analyse the homogeneous scalar field dynamics in an FLRW background. In Sec. \ref{Sec4}, we investigate scalar perturbations and their role in tachyonic amplification. In Sec. \ref{Sec5}, we discuss the possible cosmological history of hairy PBHs and present hairy PBH solutions for cosmologically consistent parameter choices. In Sec. \ref{Sec6}, we calculate the greybody factors for these hairy PBHs. Finally, we present our conclusions in Sec. \ref{Sec7}.

\section{Setup}\label{sec:setup}
We consider the following action
\begin{align}
		S = \int \rmd^4 x \sqrt{-g} \bigg[\frac{1}{2 \kappa}R - \nabla_{\alpha} \varphi^* \nabla^{\alpha} \varphi - f(\varphi^*, \varphi) \mathcal{G} - V(\varphi^*, \varphi) \bigg] \label{eq:actn}
		\end{align}
where $\kappa^2=m_P^{-2}=8\pi G$, with $m_P$ and $G$ denoting the reduced Planck mass and Newton’s gravitational constant, respectively. The GB invariant $\mathcal{G}$ is defined as
\begin{align}
    \mathcal{G}= R_{\mu \nu \alpha \beta} R^{\mu \nu \alpha \beta} - 4 R_{\mu \nu} R^{\mu \nu} + R^2\,.
\end{align}
We employ
			\begin{align}
			f(\varphi^*, \varphi) =  \frac{\lambda}{\Lambda_1^4} (\varphi^* \varphi)^2 -\frac{\alpha}{\Lambda_2^2} \varphi^* \varphi, \qquad V(\varphi^*, \varphi) =   \eta (\varphi^* \varphi)^2 \,,
            % + m^2 \varphi^*(r) \varphi(r)\,,\label{Eq:Pot}
			\end{align}
% and fix $\alpha_i,\lambda_i>0$. \sjd{[Let's keep both positive? Can we have stable hairy black holes  with either $\alpha_1, \lambda_2=0$. We can later take some of the above couplings to be zero/small if required]}
where $\lambda, \alpha$ and $\eta$ are positive dimensionless constants. The effective potential is given by
\begin{align}
    V_{\rm eff} (\varphi^*, \varphi)= f(\varphi^*, \varphi) \mathcal{G} +V(\varphi^*, \varphi)\,.\label{eq:Veff}
\end{align}
The shape of this effective potential, governed by the behaviour of the GB term, plays a crucial role in the scalar field dynamics in both black hole and FLRW spacetimes. 

This gravity action enjoys global $U(1)$ symmetry 
\begin{align}
\varphi(r) \rightarrow e^{i \chi} \varphi(r) 
\end{align}
where $\chi$ is a constant. The scales $\Lambda_1$ and $\Lambda_2$ characterize the suppression of higher-dimensional operators. Assuming that the two operators originate from the same underlying physics, we organize a single characteristic suppression scale,
\begin{align}
\Lambda_1\simeq\Lambda_2\equiv\Lambda.
\end{align}
Although additional U(1)-invariant higher-dimensional operators, such as ($\varphi^*\varphi)^n\mathcal G$ with ($n\geq3$), are generally allowed, we truncate the coupling function at quartic order since operators of higher dimension are suppressed by additional powers of $|\varphi|/\Lambda$. We also omit an explicit bare mass term, assuming that the corresponding intrinsic scalar mass is negligible compared with the curvature and self-interaction-induced effective mass scales that govern the dynamics at high energies. In principle, the symmetries of the theory also allow a nonminimal coupling of the form \(\xi |\varphi|^2R\). In the present work, we restrict the curvature-dependent scalar interactions to the coupling with the GB invariant and set \(\xi\ll1\). This choice defines the particular subclass of scalar–tensor theories considered here and allows us to isolate the effects induced by the scalar–GB coupling. We further assume that any radiatively generated Ricci-scalar coupling remains negligible over the range of scales relevant to our analysis.  The scale ($\Lambda$) introduced above need not coincide with the UV cutoff ($\Lambda_{\rm UV}$) of the complete gravitational theory. In the present analysis, we assume a hierarchy
\begin{align}
\Lambda<\Lambda_{\rm UV},
\end{align}
such that the characteristic scale associated with the scalar–GB interaction remains below the scale at which the full UV completion becomes relevant. The resulting constraints on these scales will be discussed below. 

%Thus, we have an effective potential given by
%\begin{align}
%    V_{\rm eff} (\varphi^*, \varphi)= f(\varphi^*, \varphi) \mathcal{G} +V(\varphi^*, \varphi)\label{eq:Veff}
%\end{align}
%The shape of the above potential, determined by the behavior of the GB term is crucial for the dynamics of the scalar field, both in the black hole and FLRW spacetime. We discuss the behavior in these two spacetimes below.

% \sjd{EFT discussion, bare mass term, linear $R$ term: $(\varphi^* \varphi)R$  }
%\subsection{EFT}
\subsection{Behavior in black hole spacetime}

% \cmb{
% \begin{align}
% m_{p} = \frac{1}{\sqrt{8 \pi G}} \approx 2.435 \times 10^{18} GeV \approx 4.34 \times 10^{-6} g
% \end{align}
% } 
%The above setup can lead to stable scalar hairy black hole solutions in the symmetry-broken phase, as studied in  \cite{Latosh:2023cxm}. With the metric in the black hole spacetime given by
%\begin{align}
%    ds^2 =  -A(r) dt^2 + \frac{1}{B(r)} dr^2 +r^2 (d \theta^2 +\sin^2\theta ~ d\phi^2)\,,\label{eq:BHmetric}
%\end{align}
%and we use 
%\begin{align}
%\varphi (r)= \frac{\varphi_1 (r) +i \varphi_2 (r)}{\sqrt{2}}
%\end{align}
%and set $\varphi_1 = \varphi_2$ for simplicity. For the Schwarzschild black hole of mass $M_{\rm Sch.}$, it yields $A(r)= B(r)= 1-\frac{2GM_{\rm Sch.}}{r}$. In our theory, the Schwarzschild black hole can become unstable, when \cite{1995AmJPh..63..256B}

The above setup admits stable scalar-hairy black hole solutions in the symmetry-broken phase, as demonstrated in Ref. \cite{Latosh:2023cxm}. We adopt the static, spherically symmetric metric ansatz
\begin{align}
ds^2=-A(r)dt^2+\frac{dr^2}{B(r)}
+r^2\left(d\theta^2+\sin^2\theta ~d\Phi^2\right)\,,
\label{eq:BHmetric}
\end{align}
and decompose the complex scalar field as
\begin{align}
\varphi(r)=\frac{\varphi_1(r)+i\varphi_2(r)}{\sqrt{2}}\,.
\end{align}
For simplicity, we take \(\varphi_1=\varphi_2\). For a Schwarzschild black hole of mass \(M_{\rm Sch.}\), the metric functions are
\begin{align}
A(r)=B(r)=1-\frac{2GM_{\rm Sch.}}{r}.
\end{align}
In the present theory, the effective potential (\ref{eq:Veff}) is given by 
\begin{align}
    V_{\rm eff} (r) = \frac{l (l+1)A}{r^2} +\frac{1}{2r}(A' B +BA')+ \frac{1}{2}f_{\varphi_1 \varphi_1} A \mathcal{G}\,,\label{eq:Veffr}
\end{align}
and the Schwarzschild solution becomes unstable when the following condition is satisfied \cite{1995AmJPh..63..256B}:
\begin{align}
    \int_{r_h} ^\infty \frac{V_{\rm eff} (r)}{\sqrt{A B}} dr < 0\,. \label{eq:Vbnd}
\end{align}
Here, a prime denotes differentiation with respect to $r$ and the subscript of $f$ denotes the derivative with respect to the corresponding variable. For the Schwarzschild metric, the GB term is given by
\begin{align}
    \mathcal{G}_{\rm{Sch.}} (r) = \frac{48 G^2 M_{\rm Sch.}^2}{r^6}\,.\label{eq:GBBH}
    \end{align}
For spherically symmetric perturbations ($l=0$), the bound given by Eq. \eqref{eq:Vbnd} with zero scalar field value ($\varphi_1=0$) leads to the following condition for the existence of hairy black hole solutions 
\begin{align}
  \frac{\alpha}{\Lambda_2^2} \gtrsim \frac{5}{6} G^2 M_{\rm Sch.}^2\,. \label{eq:hbhcondtn} 
\end{align}
With $\alpha \sim \mathcal{O}(1)$, the objects that develop scalar hair are therefore determined by the cut-off scale $\Lambda_2$. Defining the characteristic near-horizon energy scale as $E_{\rm BH}\equiv (G M_{\rm Sch.})^{-1}$, Eq. (\ref{eq:hbhcondtn}) gives 
\begin{align}
\Lambda_2&\lesssim \sqrt{\frac{6\alpha}{5}} \; E_{\rm BH} 
\simeq 10^{-4}\left(\frac{\alpha}{1}\right)^{1/2} \left(\frac{1~\text{g}}{M_{\rm Sch.}}\right) m_P\,. \label{eq:l2bnd}
\end{align}
Thus, when the scalar–GB suppression scale ($\Lambda_2$) is comparable to or smaller than ($E_{\rm BH}$), the sufficient condition for the instability of the symmetric Schwarzschild configuration (with $\varphi=0$) is satisfied. The subsequent growth of scalar field perturbations induces a phase transition from the Schwarzschild black hole to a hairy black hole in a symmetry-broken phase.

From Eq. \eqref{eq:l2bnd}, taking some benchmark values of black hole mass, we have
    \begin{align}
    \Lambda_2 \lesssim\begin{cases} 
    10^{13} \;\textrm{GeV}, & \textrm{for} \; \; M_{\textrm{Sch.}} \simeq 10 \textrm{g}\,,  \\
    10^{6} \;\textrm{GeV}, & \textrm{for} \; \; M_{\textrm{Sch.}} \simeq 10^8 \textrm{g} \,, \\
   10^{-1} \;\textrm{GeV}, & \textrm{for} \; \; M_{\textrm{Sch.}} \simeq 10^{15} \textrm{g}\,,  \\
    10^{-19} \; \textrm{GeV}, & \textrm{for} \; \; M_{\textrm{Sch.}} \sim M_{\odot} \simeq 10^{33} \textrm{g}\,.
    \end{cases}\label{eq:l2bmrk}
    \end{align}
As mentioned in the Introduction, we consider our theory at higher cutoff scales and extend our analysis to black hole masses within the PBH mass range. From an observational perspective, however, the objects with scalar hair typically considered in earlier studies \cite{Anderson:2016aoi, Franchini:2019npi, Anson:2019uto}, are astrophysical black holes with masses of order $M_{\rm Sch.}\sim M_{\odot}$.

In this work, we assume that the bare scalar self-interaction is negligible in the black hole spacetime. We implement this assumption by requiring the $\eta$-dependent term to remain subdominant to the GB-induced quartic interaction. This condition allows us to estimate the appropriate cutoff scale $\Lambda_1$ and the radial scale $r_B = c_B r_{\rm Sch.}$, which characterises the region within the Hubble radius where the black hole spacetime approximation remains valid. 

We first compare the magnitude of the GB-induced quartic interaction with that of the \(\eta\)-dependent term using representative values of \(M_{\rm Sch.}\) and \(\Lambda_1\), and examine how their relative importance varies with the radial scale \(r_B\). To satisfy the hairy black hole formation condition in Eq. \eqref{eq:l2bnd}, we demonstrate the case of \(\Lambda\simeq 10^{13}~\mathrm{GeV}\), which for \(M_{\rm Sch.}\simeq 1~\mathrm{g}\) gives
% Assuming $\Lambda_1=\Lambda_2\equiv\Lambda$ and $\lambda\sim\mathcal O(1)$, we find, as expected, that near the black-hole horizon the contribution of the quartic scalar–Gauss–Bonnet coupling to the effective potential dominates over that of the bare quartic self-interaction, as shown below.
\begin{align}
\frac{\mathcal{G}}{\Lambda_1^4}  = \frac{48 G^2 M_{\rm Sch.}^2}{r^6 \Lambda_1^4} \sim \begin{cases} 
10^{5} & \textrm{at} \; \; r =r_{\rm Sch.}, \\
1 & \textrm{at} \; \; r = 8  ~r_{\rm Sch.}, \\
10^{-13} & \textrm{at} \; \; r = 10^3 ~ r_{\rm Sch.}\,. 
\end{cases}
\end{align}
While for $M_{\rm Sch.} \simeq 10~\text{g}$, we get
\begin{align}
\frac{\mathcal{G}}{\Lambda_1^4}  \simeq \frac{48 G^2 M_{\rm Sch.}^2}{r^6 \Lambda_1^4} \sim \begin{cases} 
40 & \textrm{at} \; \; r = r_{\rm Sch.}, \\
  1 & \textrm{at} \; \; r = 2 ~r_{\rm Sch.}, \\
10^{-17} & \textrm{at} \; \; r = 10^3 ~ r_{\rm Sch.}\,.
\end{cases}
\end{align}
These numerical values show the rapid fall of the GB-induced quartic interaction, as we move away from the horizon. As we will see later, our allowed parameter space would be restricted to be around the black hole mass range of $1~ \text{g}$-$10 ~\text{g}$. A systematic expression for this condition, i.e. $\frac{\lambda \mathcal{G} (r=r_B)}{\Lambda_1^4}>\eta$, can be written as
%These numerical values show the rapid fall off of the GB-induced quartic interaction. If we want to keep the condition $\lambda\sim \eta\sim \mathcal{O}(1)$ until around $r \sim 10^3 \; r_{\textrm{Sch.}}$, it requires an even smaller value of $\Lambda$. Now, let us take $\Lambda\sim 10^{9} ~ \text{GeV}$ and $M_{\rm Sch.}\simeq 1~ \text{g}$, we have

%Later, we will investigate whether such lower values of $\Lambda$ are consistent in our scenario. Similar arguments hold for other values of the black hole mass. 

%At the cosmological scale, the $\eta$ term would dominate at a distance $r \gtrsim r_B$ away from the horizon, when  $\eta \sim  \frac{\lambda \mathcal{G}(r_B)} {\Lambda_1^4}$, which gives
%\begin{align}
 %  r_B \sim  \left(\frac{10^{-4} ~m_P}{ \Lambda_1}\right)^{2/3} \left(\frac{1 ~\text{g}}{M_{\rm Sch.}}\right)^{2/3}\left(\frac{
 %  \lambda
%   }{\eta}\right)^{1/6}  r_{\rm Sch.}\,, 
%\end{align}
%with $r_B > r_{\rm Sch.}$. 
%Thus, for the black hole curvature to dominate until a distance $r_B = c_B~r_{\rm Sch.}$ away from the horizon, we require 
\begin{align}
    \Lambda_1 \lesssim  10^{-4} c_B^{-3/2} \left(\frac{1 ~\text{g}}{M_{\rm Sch.}}\right) \left(\frac{
   \lambda
   }{\eta}\right)^{1/4} m_P \,,\label{eq:L1bndBH} 
\end{align}
with $c_B>1$. We analyse whether these parameters, leading to consistent hairy black hole solutions with a subdominant $\eta$ term in black hole spacetime, remain valid in a cosmological history. 

% Beyond $r_B$, the cosmological FLRW background becomes dominant. As the Gauss–Bonnet contribution decreases during the post-inflationary evolution, the bare self-interaction eventually becomes dynamically relevant. 
%  We analyse whether these solutions remain valid in a cosmological history.
\subsection{Behavior in cosmological background}
\begin{figure}[h]
\includegraphics[scale=0.28]{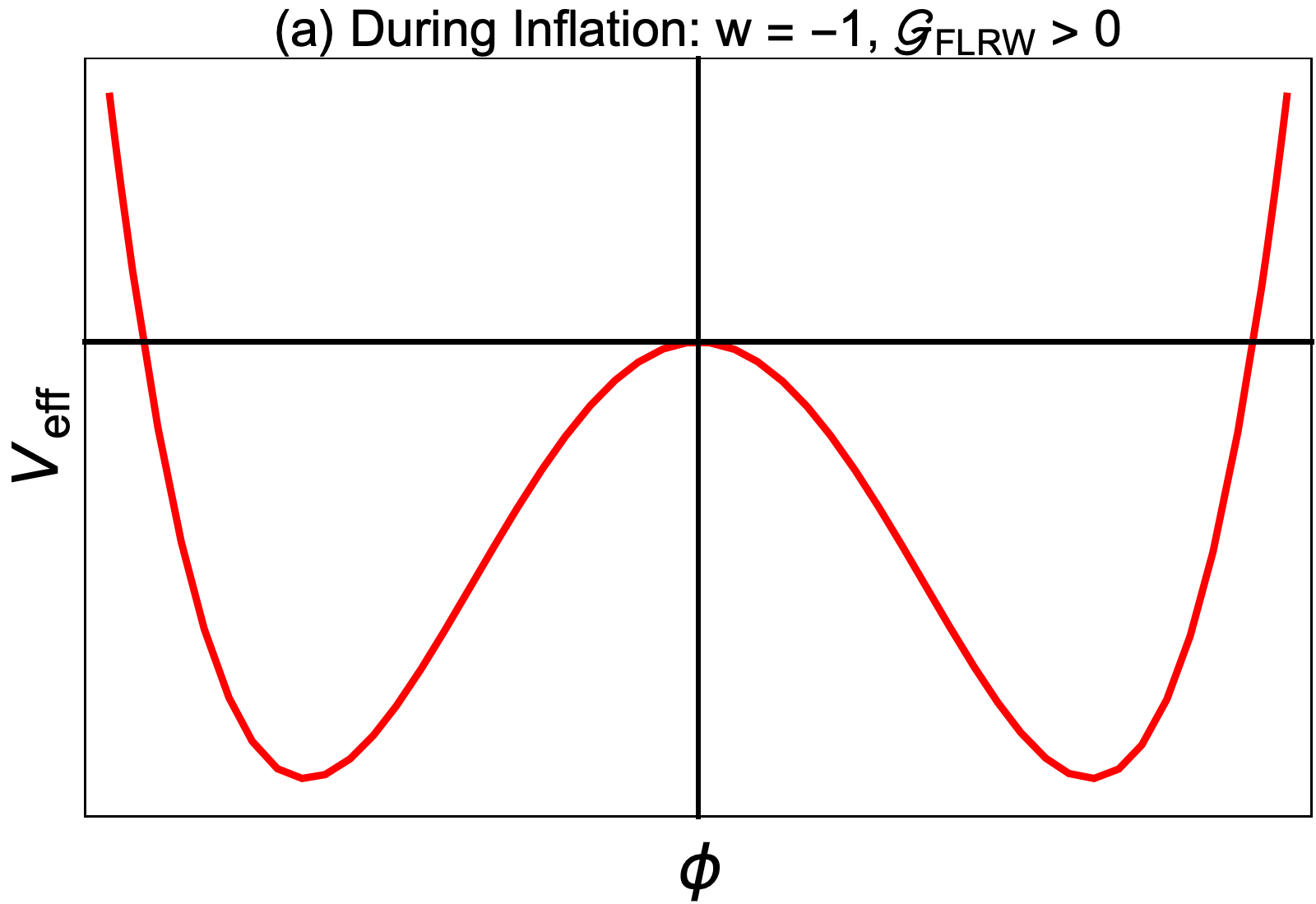} ~\includegraphics[scale=0.28]{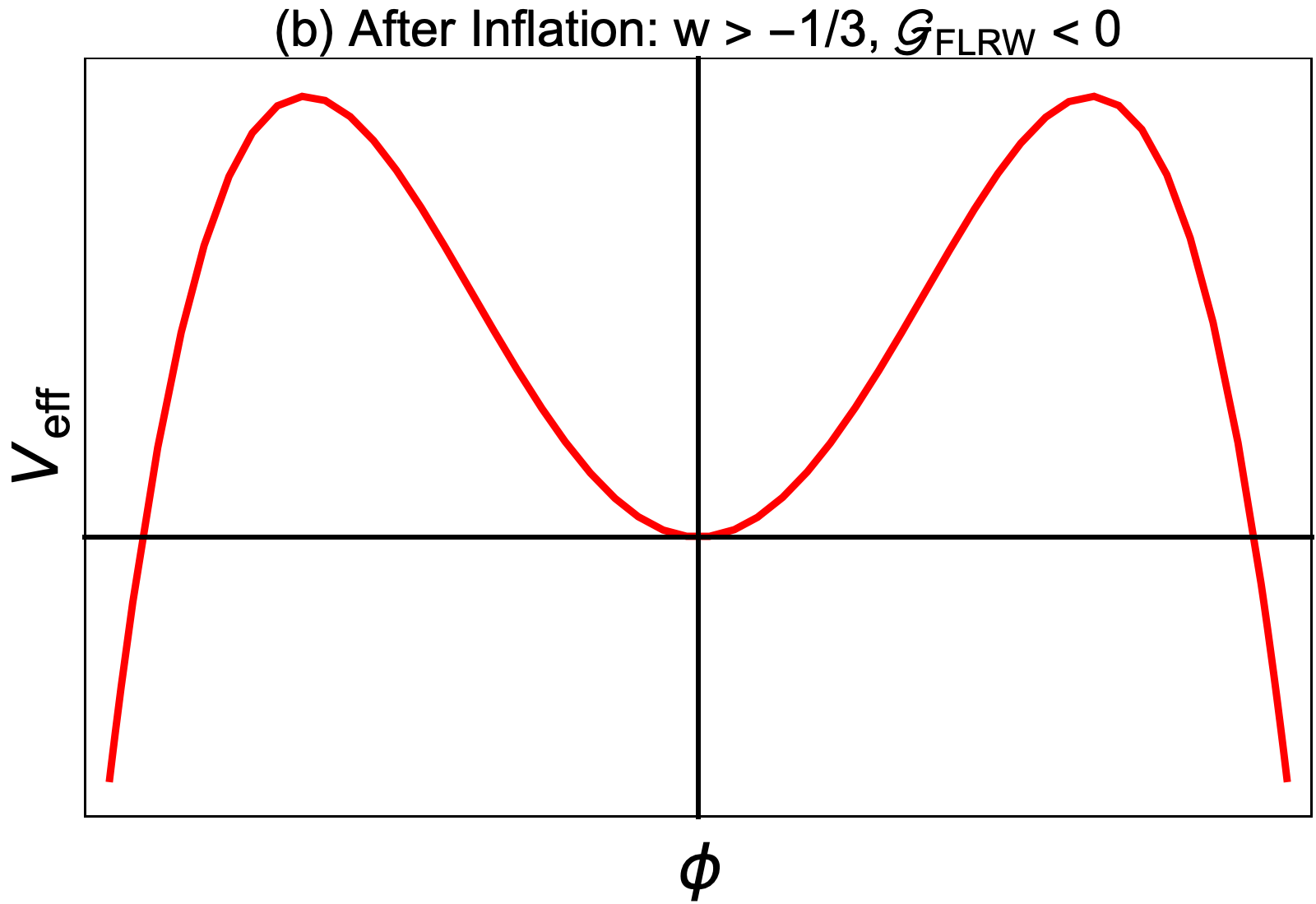}~
\includegraphics[scale=0.28]{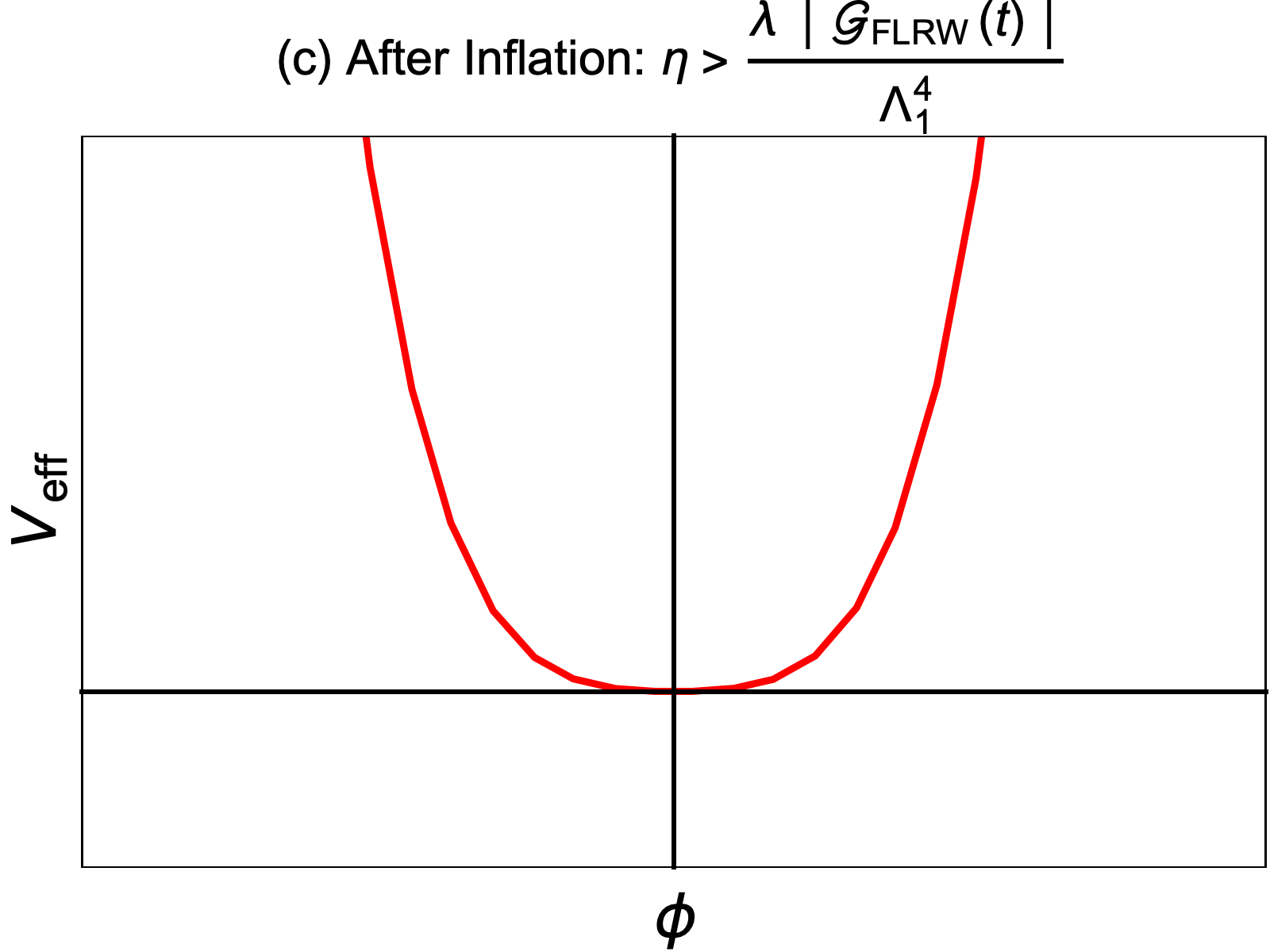}\\
\begin{center}
 \includegraphics[scale=0.5]{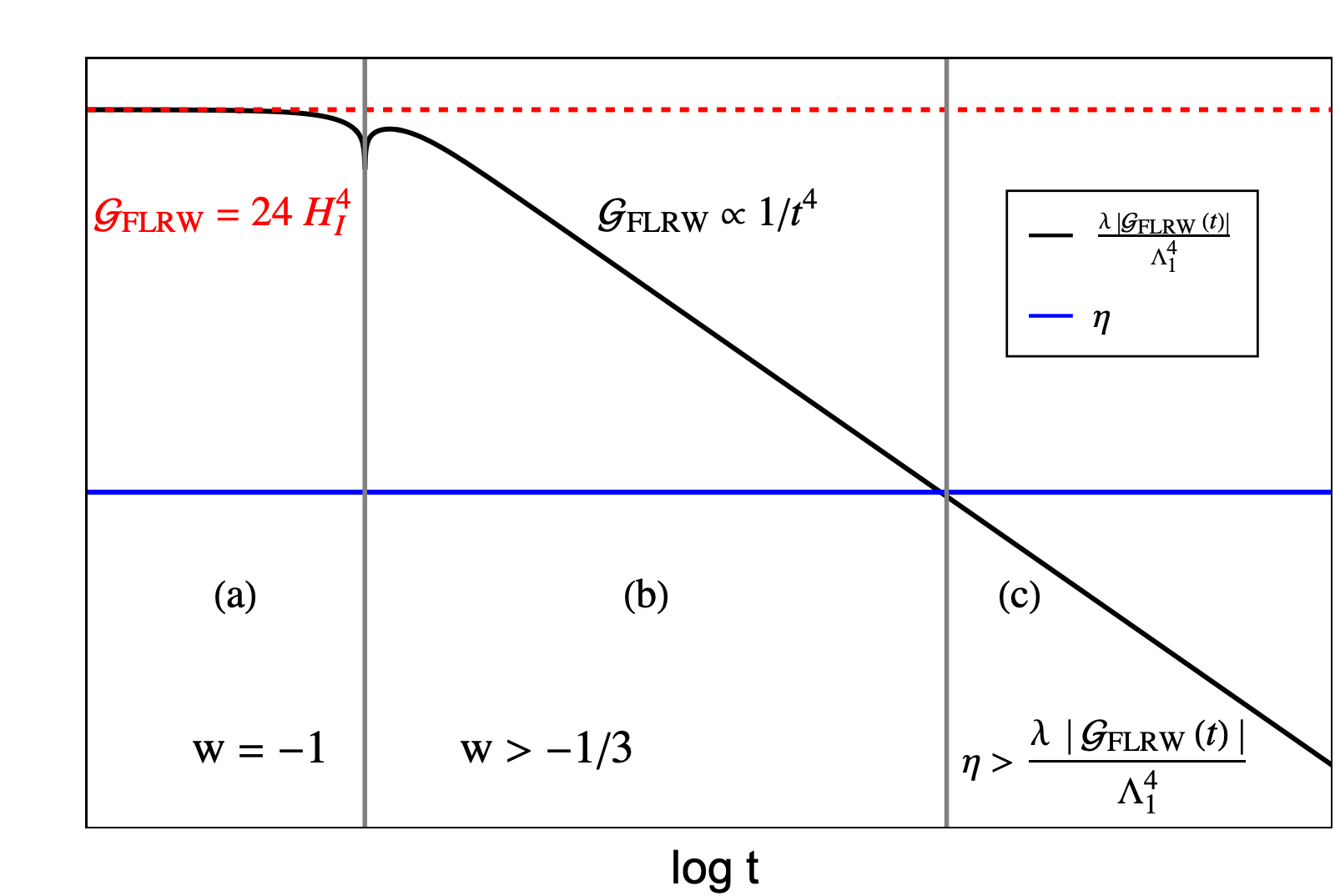}   
\end{center}
\caption{\textit{Upper panel:} Schematic showing the shape of the effective potential $V_{\rm eff}$ (Eq. \eqref{eq:Veff}) in different cosmological epochs: (a) during inflation ($w=-1$), (b) after inflation ($w>-1/3$), (c) after $\eta > \frac{\lambda |\mathcal{G}_{\rm FLRW}(t)|^{4}}{\Lambda_1^4 }$. \textit{Lower panel}: Schematic of the evolution of $\frac{\lambda |\mathcal{G}_{\rm FLRW}(t)|^4}{\Lambda_1^4}$. The red dashed line indicates its maximum value, which is during inflation. The blue line denotes $\eta$.} 
\label{fig:potFLRW}  
\end{figure} 
The background FLRW metric is given by
\begin{align}
    ds^2= -dt^2 + a(t)^2 \delta_{ij}dx^i dx^j\,,
\end{align}
where $a(t)$ indicates the scale factor. The Gauss-Bonnet invariant in this background metric is given by
\begin{align}
    \mathcal{G}_{\rm{FLRW}}&= 24 H^2 \frac{\ddot{a}}{a}\nonumber\\
    &=\begin{cases}
    24H_I^4~\text{for}~w=-1\,,\\ 
        -\frac{64(1+3w)}{27(1+w)^4t^4} ~\text{for}~w\neq-1 \,.\label{eq:GBFLRW} 
    \end{cases}
\end{align}
where $w$ denotes the equation of state parameter, and $H_I \simeq 10^{13}$ GeV \cite{Planck:2018jri}, indicates the typical Hubble scale during inflation. In terms of $w$, $\mathcal{G}_{\rm{FLRW}}$ is positive for $-1\leq w\leq -1/3$ and becomes negative for $w>-1/3$. Thus, for accelerated expansion of the Universe ($\ddot{a}>0$)  during inflation and the present-day dark energy epoch, we have $\mathcal{G}_{\rm{FLRW}}>0$. On the other hand, in the radiation-dominated (RD) ($w=1/3$) and matter-dominated ($w=0$) Universe, with decelerated expansion ($\ddot{a}<0$), we have $\mathcal{G}_{\rm{FLRW}}<0$. 

The upper panel of Fig. \ref{fig:potFLRW} shows the schematic of the shape of the effective potential $V_{\rm eff}$, given by Eq. \eqref{eq:Veff}, in different cosmological epochs. We discuss these three essential epochs below\footnote{See Refs. \cite{Liang:2019fkj, Aldabergenov:2025oys, Krasnov:2025ssc, Laverda:2026slq} for related studies of spontaneous symmetry breaking in the cosmological background, in the presence of GB coupling.}:

(a) \textit{During inflation:} With positive $\mathcal{G}_{\rm FLRW}$ and $\eta$, the symmetry of the potential remains spontaneously broken during this epoch. The vacuum expectation value (vev) in the symmetry broken phase is given by
\begin{align}
 |<0|\varphi|0>| = v_{\rm \varphi}=\sqrt{\frac{\alpha \mathcal{G}/\Lambda_2^2}{2 (\frac{\lambda\mathcal{G}}{\Lambda_1^4}+\eta)}}\,.\label{eq:vev1}
\end{align}
Let us again compare the value of the GB $\lambda$ term with the quartic $\eta$ term, but this time considering the FLRW background for the calculation of $\mathcal{G}$. During inflation, we have
\begin{align}
    \frac {\lambda}{\Lambda_1^4} |\mathcal{G}_{\rm FLRW} (t)| = 24\lambda \frac {H_I^4}{\Lambda_1^4}\,. \label{eq:infcompare}
\end{align}
Now, in order for the hairy black hole solution to remain consistent, the coefficient  $\frac {\lambda}{\Lambda_1^4}$ should satisfy the lower bound given by Eq. \eqref{eq:L1bndBH}, i.e.
\begin{align}
    \frac{\lambda}{\Lambda_1^4} \gtrsim 10^{16}~\eta~ c_{B}^6 \left(\frac{M_{\rm Sch.}}{1~\text{g}}\right)^4 m_P^{-4}\,.
\end{align}
Hence, Eq. \eqref{eq:infcompare} leads to
\begin{align}
     \frac {\lambda}{\Lambda_1^4} |\mathcal{G}_{\rm FLRW} (t)| \gtrsim 2.4 \times 10^{17} ~\eta ~c_{B}^6 \left(\frac{M_{\rm Sch.}}{1~\text{g}}\right)^4 \left(\frac{H_I}{m_P}\right)^4\,.\label{eq:GBdom}
\end{align}
The GB contribution is  therefore much larger than $\eta$ 
for any feasible black hole, of mass at least a few grams\footnote{The typical lower bound on black hole mass is discussed later in Section \ref{Sec5} (cf. Eq. \eqref{Eq:PBHlowbnd}).}. For eg., $c_B=100$ gives the above value to be around  $10^{8} \eta$ for $M_{\rm Sch.} =1~\text{g}$.

Thus, just like in the black hole spacetime, the $\eta$ term can be neglected for the FLRW spacetime during inflation as well, and the vev given by Eq. \eqref{eq:vev1} is simplified to be
\begin{align}
    v_{\varphi} \simeq \sqrt{\frac{\alpha}{2 \lambda}} \frac{\Lambda_1^2}{\Lambda_2}\,. \label{eq:vev2}
\end{align}

(b)\textit{After inflation, $\mathcal{G}_{\rm FLRW}<0$:} After the end of inflation, the Universe enters the decelerating phase and the GB induced quartic term $\mathcal{G}_{\rm FLRW}$ becomes negative as discussed earlier. Since this term is larger than the quartic $\eta$ term as just established above via Eq. \eqref{eq:GBdom}, the effective potential becomes unbounded in the presence of the GB induced negative quadratic $\alpha$ term. In the absence of other stabilising positive terms in the potential, which we don't consider here to be effective at these energy scales, the potential remains unbounded until the negative GB term becomes small enough, which we discuss next. 

(c) \textit{After inflation, $\eta >\frac{\lambda |\mathcal{G}_{\rm FLRW}|}{\Lambda_1^4}$:} During the decelerated phase, the GB induced quartic contribution decreases as $t^{-4}$ (cf. Eq. \eqref{eq:vev1}). The bare self-interaction therefore eventually dominates and restores the symmetry of the scalar potential,  with a global minimum now at $|\varphi|=0$. The symmetry-restoration time is given as
\begin{align}
t_{\rm res.}
\simeq
\frac{1}{\Lambda_1}
\left[
\frac{64\lambda |1+3w|}
{27\eta(1+w)^4}
\right]^{1/4}\,.
\end{align}

 The lower panel of Fig. \ref{fig:potFLRW}  shows a schematic of the evolution of the GB term: $\frac{\lambda |\mathcal{G}_{\rm FLRW}(t)|^{4}}{\Lambda_1^4}$. It has the maximum value during inflation, when $\mathcal{G}_{\rm FLRW} (t)$ is largest, indicated by the red-dashed line. After inflation, $\mathcal{G}_{\rm FLRW} (t)$ becomes negative and decreases rapidly as $1/t^{4}$. As a result, $\eta$ (blue line), starts to dominate in region (c)\footnote{During the dark-energy dominated epoch at present time, the symmetry can be spontaneously broken again because of positive $\mathcal{G_{\rm FLRW}}$. However, during such late times, the GB term has dropped enough and is vanishingly small, and hence the symmetry of the vacuum still effectively remains restored, close to zero.}. The unbounded phase at $t\lesssim t_{\rm res.}$ in region (b), can make the scalar field unstable in the FLRW background. Hence, in the next section, we analyse the dynamics of the scalar field in details throughout the cosmic history.

\begin{figure}
    \includegraphics[scale=0.4]{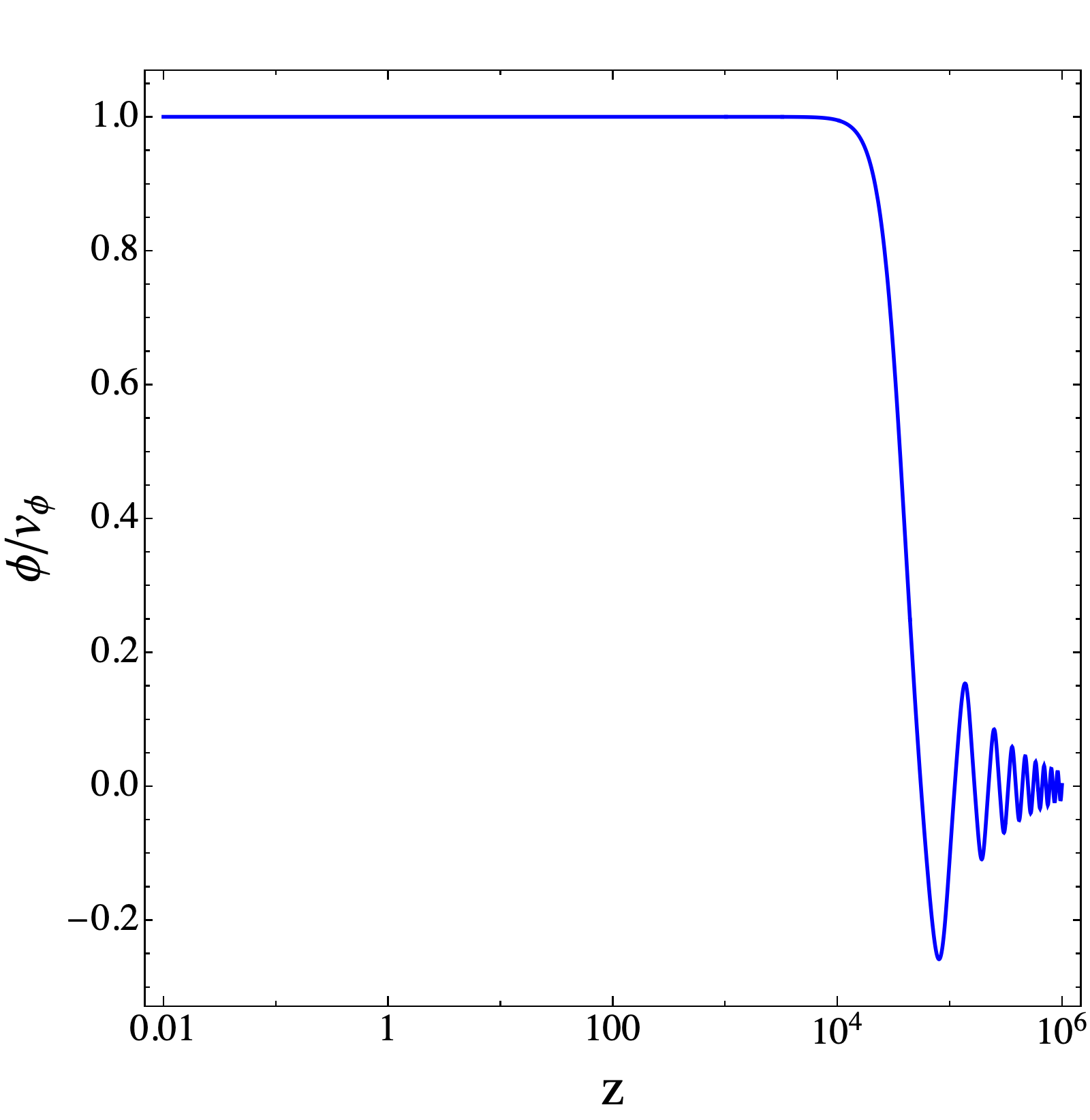} ~~
    \includegraphics[scale=0.4]{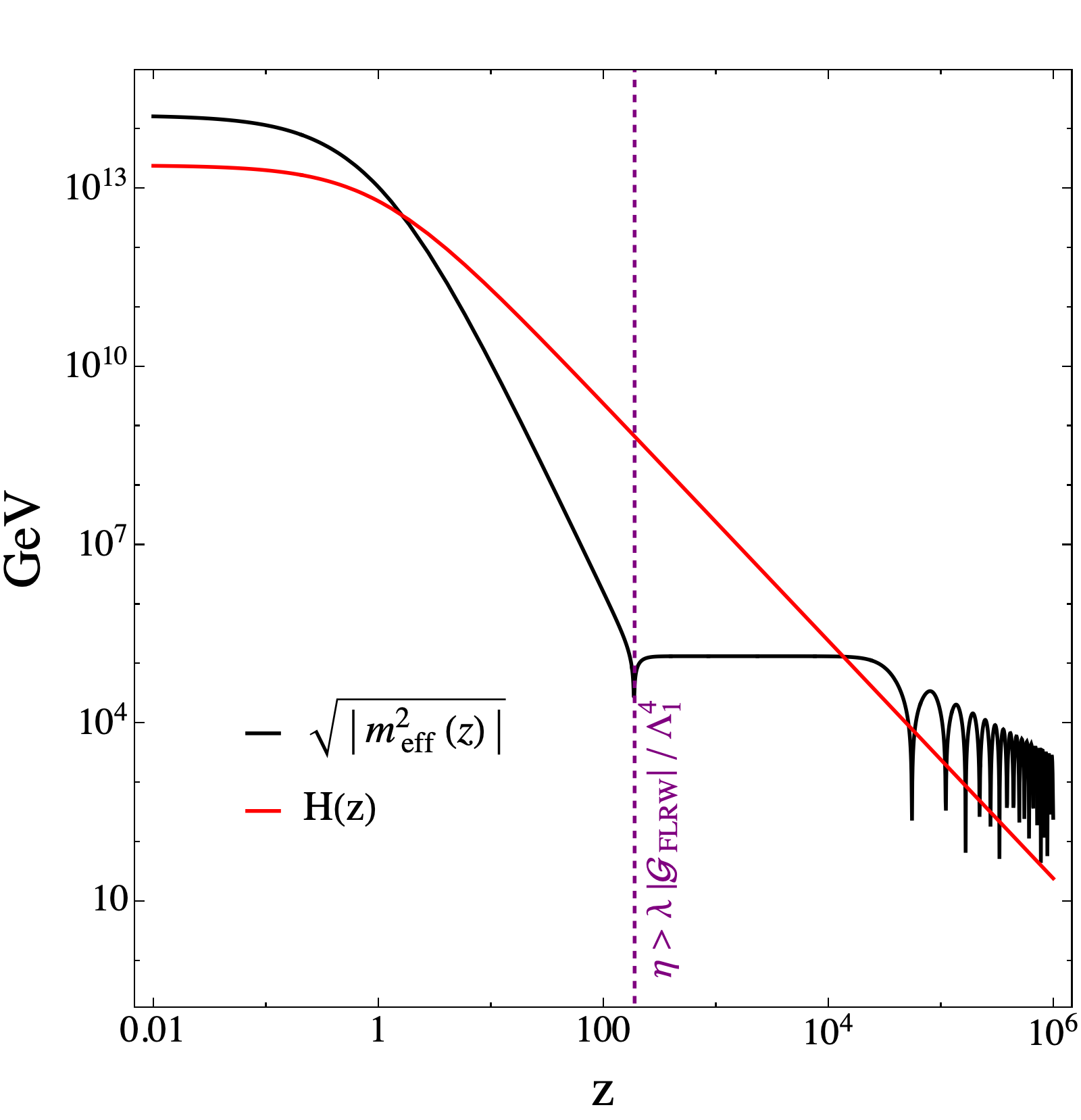}
\caption{\textit{Left panel}: Evolution of the scalar field considering it to be at $\phi=v_{\phi}$, with $\dot{\phi} =0$ just after inflation at $z=0$. \textit{Right panel}: Evolution of the effective mass given by $\sqrt{|m^2_{\rm eff} (z)|}= \sqrt{\frac{\partial^2 V}{\partial \phi ^2}}$, along with the Hubble rate $H(z)$. The vertical dashed line indicates the epoch when the GB term has decreased enough such that the $\eta$ term dominates, restoring the symmetry. Note that the field oscillates after the Hubble rate has dropped below the constant $|m_{\rm eff} (z)|$.} 
\label{fig:hmo}  
\end{figure} 
\section{Scalar field dynamics in FLRW cosmology}\label{Sec3}

In the FLRW spacetime, the global U(1) symmetry of the vacuum remains spontaneously broken during inflation. Writing $\varphi = \frac{\phi}{\sqrt{2}}e^{i\theta}$, where $\phi$ indicates the radial component of the complex scalar field, the effective mass square of the field $\phi$ in the symmetry-broken phase during inflation is given as\footnote{In the presence of an additional linear $R$ term of the form $\xi |\varphi|^2R$, there would be an additional contribution to the effective mass square,  which is of the order $\xi H_I^2$. However, it is negligible for $\xi \lesssim \frac{\alpha H_I^2}{\Lambda_2^2}$.} 
%\sjd{[I think let's not write it by expanding around v here? Otherwise $V$ carries other $v$-dependent terms and hence also EOM would have such terms, and might change the equations below, although final results should be same. I think if we define $v$ of the complex field $\Phi$ as in Eq. 21, and write  $\varphi = \frac{\phi}{\sqrt{2}}e^{i\theta}$, rest would remain the same. $|\Phi|=v$ would imply $\phi = \sqrt{2} v$, and other  calculations like $V''$ and EOM would remain the same.]}
\begin{align}
  m_{\rm eff}^2&=  V''(\phi=v_{\phi})\nonumber\\ &\simeq\frac{2\alpha \mathcal{G}_{\rm{FLRW}}}{\Lambda_2^2}  \nonumber\\
  &\simeq 48 \frac{H_I^4}{\Lambda_2^2} \alpha \,.\label{eq:meffinf}
\end{align}
We consider the field to be sufficiently heavy during inflation, such that the field is located at the minimum and any isocurvature perturbations are suppressed. Thus, we require
\begin{align}
    m_{\rm eff}> H_{I} \implies \Lambda_2 \lesssim 4 \sqrt{3 \alpha} ~H_I\,. \label{eq:iso1}
\end{align}
% Since we want to generate hairy black hole solutions, we want $\Lambda_2$ to follow the condition given by Eq. \eqref{eq:hbhcondtn}, which gives us the following lower bound on the Schwarzschild black hole mass that can develop scalar hair
% \begin{align}
%     M_{\rm Sch.} \gtrsim 1.3 ~\text{g}\,. \label{eq:iso2}
% \end{align}

Now, after the end of inflation, $w$ transforms to the value $1/3$ corresponding to a RD Universe, during the reheating process, the details of which depend on the reheating model.  Regardless of the reheating model, once $w$ becomes larger than $-1/3$, $\mathcal{G}_{\rm FLRW}$ changes sign (cf. Eq. \eqref{eq:GBFLRW}). Thus, the potential flips and becomes unbounded, as discussed earlier, and the effective mass square becomes negative, given by
\begin{align}
    m_{\rm eff}^2 \simeq -   \frac{48\alpha H(t)^4}{\Lambda_2^2}\,, 
\end{align}
the magnitude of which decreases with time as $1/t^4$. Because of this tachyonic mass at the top of the potential, the quantum fluctuations of the  scalar field can grow rapidly \cite{Felder:2001kt}, leading to an instability that can destroy the standard cosmological background. This unstable tachyonic phase can last until the positive $\eta$ term dominates, thereby restoring the symmetry (cf. Fig. \ref{fig:potFLRW}).

% and the symmetry of the potential is restored.  The field then moves to the new minima at zero. A large value of the  vev $v_{\rm cosmo}$ in the symmetry broken phase, i.e. $v_{\rm cosmo}\gtrsim M_{P}$, in addition to being unnatural, would also lead to a large energy in the scalar field after symmetry restoration, disturbing the standard cosmological picture. This can be used to put an upper bound on the BH mass that can lead to hairy black hole solutions, as we show below.
Let us now analyse the dynamics of the scalar field in detail. Varying the action in Eq. \eqref{eq:actn} with respect to the field, gives the Klein-Gordon equation of motion in the FLRW background as 
 \begin{align}
     \ddot{\phi} + 3H \dot{\phi} -\frac{1}{a^2} \nabla ^2 \phi + V' (\phi) =0\,,
 \end{align}
% \begin{align}
%     \frac{3}{8 \pi}M_P^2H^2 = \rho_{\varphi} +\rho_{\rm GB} +\rho_{w}\,,
% \end{align}
% where
% \begin{align}
%    & \rho_{\varphi} = \frac12 \dot{\varphi}^2 -V(\varphi, \varphi^*)\,,\\
%     & \rho_{GB}= 24 H^3 \dot{f}\,,
% \end{align}
% while $\rho_w$ corresponds to the energy density of the background fluid satisfying the continuity equation
% \begin{align}
%     \dot{\rho}_w+3H\rho_w(1+w)=0.
% \end{align}
% As mentioned above, we need to satisfy $\rho_{\varphi}, \rho_{\rm GB} \ll  \frac{3}{8 \pi}M_P^2H^2$, in order to recover the standard cosmology. Considering the maximum energy density $\rho_{\phi}\sim V(|\varphi|=v_{\rm cosmo})$, we have the condition
% \begin{align}
%    &\Omega_{\phi}= \frac{8 \pi \rho_{\phi}}{3M_P^2 H^2}\lesssim1 \nonumber\\ 
%   & \implies \frac{8 \pi}{12} \frac{\alpha_1^2 \mathcal{G}_{\rm cosmo}^2}{\lambda_2 M_{P}^2 H^2}\lesssim1 \nonumber\\
%   &\implies M \lesssim  \left( \frac{\lambda_2}{1}\right) \left(\frac{\alpha_{\rm Sch.}}{\alpha_1}\right)^2 ~ 8.8 \times 10^{3}~  g\,,
% \end{align}
% where we have used Eq. \eqref{eq:GBFLRW} for $\mathcal{G}_{\rm cosmo}$ and Eq. \eqref{eq:alphbnd1}, such that we can have hairy black hole solutions at small scales. Additionally, we have
% \begin{align}
%     & \Omega_{\rm GB}= \frac{8 \pi \rho_{\rm GB}}{3M_P^2 H^2}\lesssim1 \nonumber\\ 
%    & \implies \frac{64 \pi H \dot{f} }{M_P^2}\lesssim1 
% \end{align}
% \sjd{[Need to check (maybe numerically) whether the above condition holds and lead to similar bound as in Eq. 22]}
% ...

% Thus,
where $\; \dot{} \;$ indicates the derivative with respect to $t$. It is well known that the dynamics of the scalar field in such a scenario is governed by the perturbations generated by quantum fluctuations, rather than the homogeneous classical background \cite{Felder:2001kt, Felder:2000hj, Bettoni:2021zhq, Laverda:2023uqv,Laverda:2026slq}. Before diving into the details of perturbations in Section \ref{Sec4}, here we provide some insight into the homogeneous scalar field background for completeness, assuming the initial conditions: $\phi=v_{\phi}$,  $\dot{\phi} = 0$, just after the flip of the potential after inflation.

We work with the following dimensionless variables \cite{Laverda:2026slq}
\begin{align}
    Y= \frac{a \phi}{a_* H_I}\,,~~~~~~~~z = a_*  H_I \tau\,,~~~~~~~~ \vec{y} =a_* H_I \vec{x}\,, \label{eq:dimlspar}
\end{align}
where $\tau$ is the conformal time defined as $d\tau = dt /a$ and `$*$' indicates a reference point chosen to be the end of inflation. The scale factor and the Hubble term are parametrised as
\begin{align}
    a(z)= a_* (1+z) \,,~~~~~H(z) = \frac{H_I}{(1+z)^2}\,.
\end{align}
We assume an instantaneous reheating such that the Universe is RD after inflation\footnote{Such an instantaneous reheating can be realised, for eg., in  some preheating models \cite{Rubio:2019ypq, Dux:2022kuk}. However, in general, reheating is model-dependent and can be a prolonged process, and we can have a different equation of state $w$ after inflation; see \cite{Barman:2025lvk} for a recent review. However, our results are not expected to depend strongly on the details of reheating, since $w>-1/3$ for any typical reheating process and $\mathcal{G_{\rm FLRW}}$ changes sign.}, starting with $z=0$, and we choose $a_*=1$. Under the above transformations and parameterisations, the equation of motion of the homogeneous mode can be written as 
\begin{align}
    Y''(z) + \frac{\partial V(Y)}{\partial Y} \frac{a(z)^4}{H_I^4}- \frac{a''(z)}{a(z)}Y(z)= 0\,,    
    % \cmb{\qquad Y''(z) + \frac{\partial V(Y)}{\partial Y} \frac{a(z)^2}{H_I^2}- \frac{a''(z)}{a(z)}Y(z)= 0}
\end{align}
where $'$ indicates derivatives w.r.t. $z$, and the last term vanishes in the RD universe. In the above, the potential is parametrised as
\begin{align}
    V(Y)= \frac{H_I^4}{a(z)^4}\left(\frac\eta4 Y(z)^4+\frac{\lambda}{4\Lambda_1^4}\mathcal{G}(z)Y(z)^4-\frac{a(z)^2}{H_I^2}\frac{\alpha}{2\Lambda_2^2}\mathcal{G}(z)Y(z)^2\right)\,.
\end{align}

If the growth of fluctuations is not considered, the field stays at the maximum of the potential after inflation, which is shown in the left panel of Fig. \ref{fig:hmo}. The right panel of Fig. \ref{fig:hmo} shows the evolution of $\sqrt{|m^2_{\rm eff}(z)|}= \sqrt{|\frac{\partial^2 V}{\partial \phi ^2}|}$ and $H(z)$. The effective mass $m_{\rm eff} (z)$ becomes constant and positive once the $\eta$ term dominates over the GB term, since the latter decreases with time. The field remains stuck at the symmetry restored potential (cf. Fig. \ref{fig:potFLRW}), until the friction from the Hubble term $H(z)$ becomes less than $m_{\rm eff}(z)$.   After this, the field moves and oscillates around zero, as shown in the left panel plot. However, this homogeneous approximation will break down as gradients develop because of the tachyonic growth of the fluctuations in the unbounded potential. We investigate this in the next section.

\section{Perturbations of the scalar field in FLRW cosmology}\label{Sec4}
Perturbing the scalar field as
\begin{align}
    \phi (\vec{x},t) = \phi (t) + \delta \phi (\vec{x},t) \,,
\end{align}
the equation of motion of the perturbations  is given as
\begin{align}
    \delta \ddot{\phi} + 3 H \delta \dot{\phi} -\frac{1}{a^2} \nabla ^2 \delta \phi+V''(\phi)\delta\phi =0\,,\label{eq:eompert}
\end{align}
The scalar perturbations can be expanded into Fourier modes as
\begin{align}
 \delta \phi(t,x) \sim \int \frac{\rmd^3\vec{k}}{(2 \pi)^3} \, \delta \phi_k (t) e^{i  \vec{k}.\vec{x}}\,.
\end{align}
The Fourier modes $\delta \phi_k$ obey the  following equation of motion
\begin{align}
     \delta \ddot{\phi_k} + 3 H \delta \dot{\phi_k} +\left(\frac{k^2}{a^2}+V''(\phi)\right)\delta\phi_k =0\,.
\end{align}
 The fluctuations of the field can be promoted to quantum operators, which, using the variables defined in Eq. \eqref{eq:dimlspar}, are given by
\begin{align}
   \delta\hat{Y}_{\kappa} (z)= f_{\kappa} (z) \hat{a}_{\kappa} +  f_{\kappa}^{*} (z) \hat{a}_{-\kappa}^{\dagger}\,.
\end{align}
The dimensionless comoving wavenumber is defined as
\begin{align}
    \kappa = \frac{|\vec{k}|}{a_* H_I}\,,
\end{align}
where $\hat{a}_{-\kappa}^{\dagger}$ and $\hat{a}_{\kappa}$ are the creation and annihilation operators satisfying the canonical commutation relations $[\hat{a}_{\kappa}, \hat{a}_{\kappa \cmb{'}}^{\dagger}]= (2 \pi)^3 \delta^3 (\kappa - \kappa')$. The Fourier modes $f_{\kappa}$ obey the equation of motion
\begin{align}
    f_{\kappa}''(z) + \omega_\kappa^2 (z)f_{\kappa} (z) =0\,,
\end{align}
where
\begin{align}
    \omega_\kappa^2 (z) = \kappa^2 +  M^2_{\rm eff} (z)\,,~~~~M^2_{\rm eff} (z)= 3 \left(\frac{\lambda}{\Lambda_1^4} \mathcal{G}_{\rm FLRW} (z)+\eta\right) Y(z)^2 - \frac{a(z)^2}{H_I^2} \frac{\alpha}{\Lambda_2^2}  \mathcal{G}_{\rm FLRW} (z)-\frac{a''(z)}{a(z)}\,. 
\end{align}
Note again that the last term in  $M^2_{\rm eff} (z)$ is zero in the RD Universe. 

 \textbf{Solution during inflation}: The solution of the mode functions in quasi de-Sitter space during inflation can be obtained using the Bunch-Davies vacuum initial conditions. At the end of inflation, the solutions are given as \cite{Riotto:2002yw, Opferkuch:2019zbd} 
\begin{align}
    f_{\kappa}  (z=0) \simeq \sqrt{\frac{\pi }{4 }}  \exp \left(\frac{1}{4} i \pi  (2 \nu +1)\right) \mathcal{H}_{\nu}^{(1)}(\kappa)\,,\label{eq:fkinf}
\end{align}
where $\nu=\sqrt{\frac{9}{4}-\frac{V''(\phi=v_{\phi})}{H_I^2}}$ and $\mathcal{H}_{\nu}^{(1)}(\kappa)$ indicates the Hankel function of the first kind. The variance of the field is defined as
\begin{align}
    \sigma_{\rm var}  (z)=\sqrt{<\delta \hat{\phi}^2>}= H_I\sqrt{\int_0^{\kappa} \frac{d^3\kappa}{(2 \pi)^3} |f_\kappa|^2}\,.
\end{align}
At the end of inflation, we have
\begin{align}
    \sigma_{\rm var}^{\rm end}  = H_I\sqrt{\int_0^{\kappa_I} \frac{d^3\kappa}{(2 \pi)^3} |f_\kappa(z=0)|^2}\,,\label{eq:sgmainf}
\end{align}
where $\kappa_I$ refers to momentum modes exiting the horizon just at the end of inflation. In the limit $\kappa \ll1$ and $\frac{V'' (\phi=v_{\phi})}{H_I^2}\gg1$, the solution given by Eq. \eqref{eq:fkinf} can be approximated to calculate the variance as \cite{Riotto:2002yw, Opferkuch:2019zbd}
\begin{align}
   \sigma_{\rm var}^{\rm end} \simeq \frac{H_I}{2 \pi} \sqrt{\frac{H_I}{3 \sqrt{V''(\phi=v_{\phi})}}}\,.
\end{align}
We verified that the above estimate agrees well with the result of the numerical integration in Eq. \eqref{eq:sgmainf}.

Now, let us estimate the value of this variance in our scenario. Using the value of $V''(\phi=v_{\phi})$ from Eq. \eqref{eq:meffinf}, we have
\begin{align}
    \sigma_{\rm var}^{\rm end} \sim 0.2 \frac{H_I}{2 \pi}\sqrt{\frac{\Lambda_2}{\sqrt{\alpha}H_I}}\,.
\end{align}
Comparing the above with the background $\phi =v_{\phi}$, 
\begin{align}
    \frac{\sigma_{\rm var}^{\rm end}}{v_{\phi}} \sim 3.2 \times 10^{-2} 
    \frac{\lambda^{1/2}}{\alpha^{3/4}} \frac{H_I^{1/2}}{\Lambda_1^2 \Lambda_2^{-3/2}}\,. \label{eq:variancecd}
\end{align}
Considering $\Lambda_1\sim \Lambda_2 = \Lambda$, we thus get
\begin{align}
    \frac{\sigma_{\rm var}^{\rm end}}{v_{\phi}} \sim 
    \frac{\lambda^{1/2}}{\alpha^{3/4}} \left(\frac{ 10^{-3}H_I}{\Lambda}\right)^{1/2}\,.
\end{align}
In order for the field to be localised around the background vev, we need to satisfy $\sigma_{\rm var}^{\rm end} \lesssim v_{\phi}$, which gives a constraint on the cut-off scale $\Lambda$ as 
\begin{align}
    \frac{\Lambda}{m_P}> 2 \times 10^{-8} \frac{\lambda}{\alpha^{3/2}}\,. \label{eq:sigmavbnd}
\end{align}
\begin{figure}[H]
\begin{center}
    \includegraphics[scale=0.37]{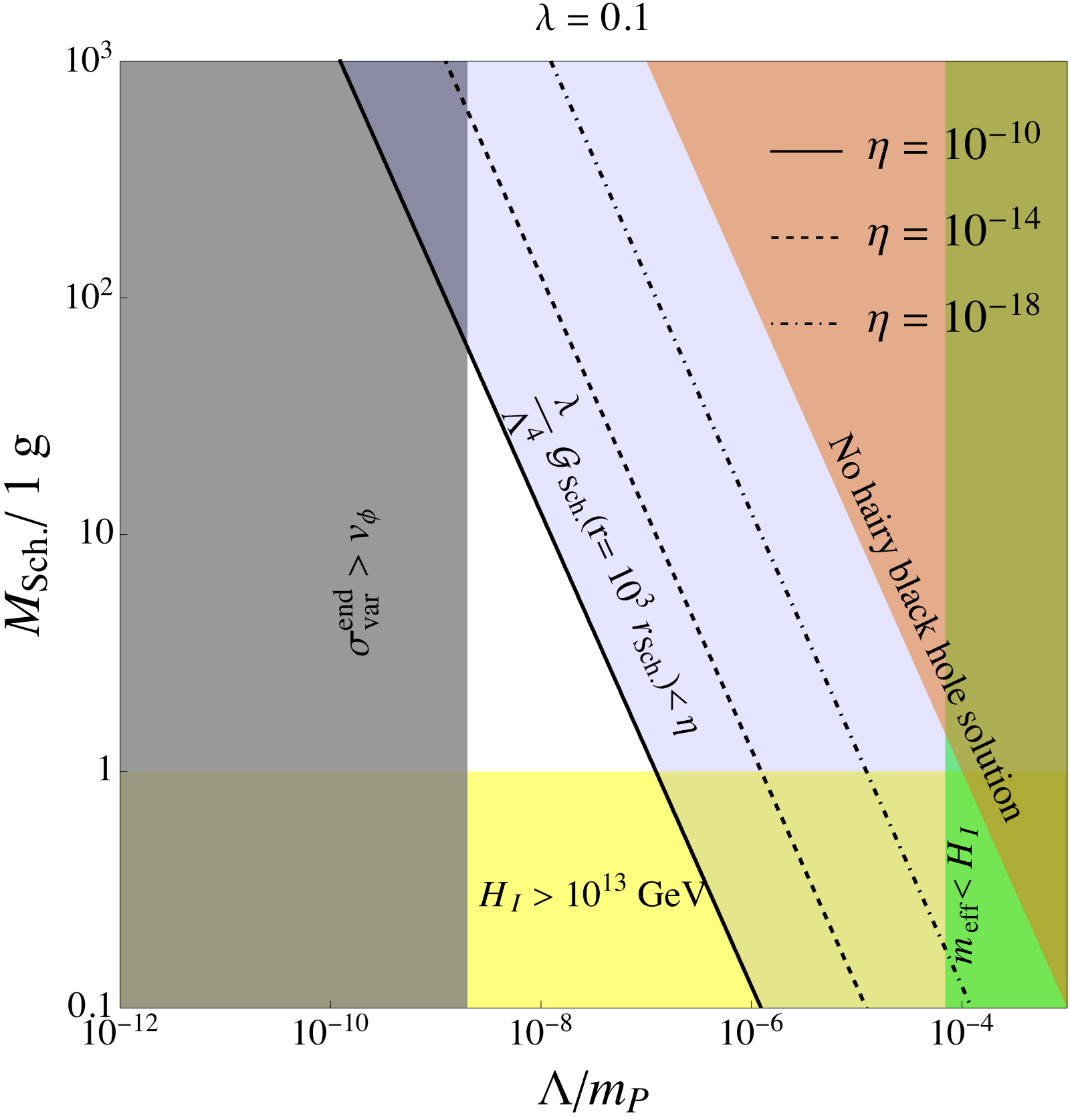} 
    \includegraphics[scale=0.37]{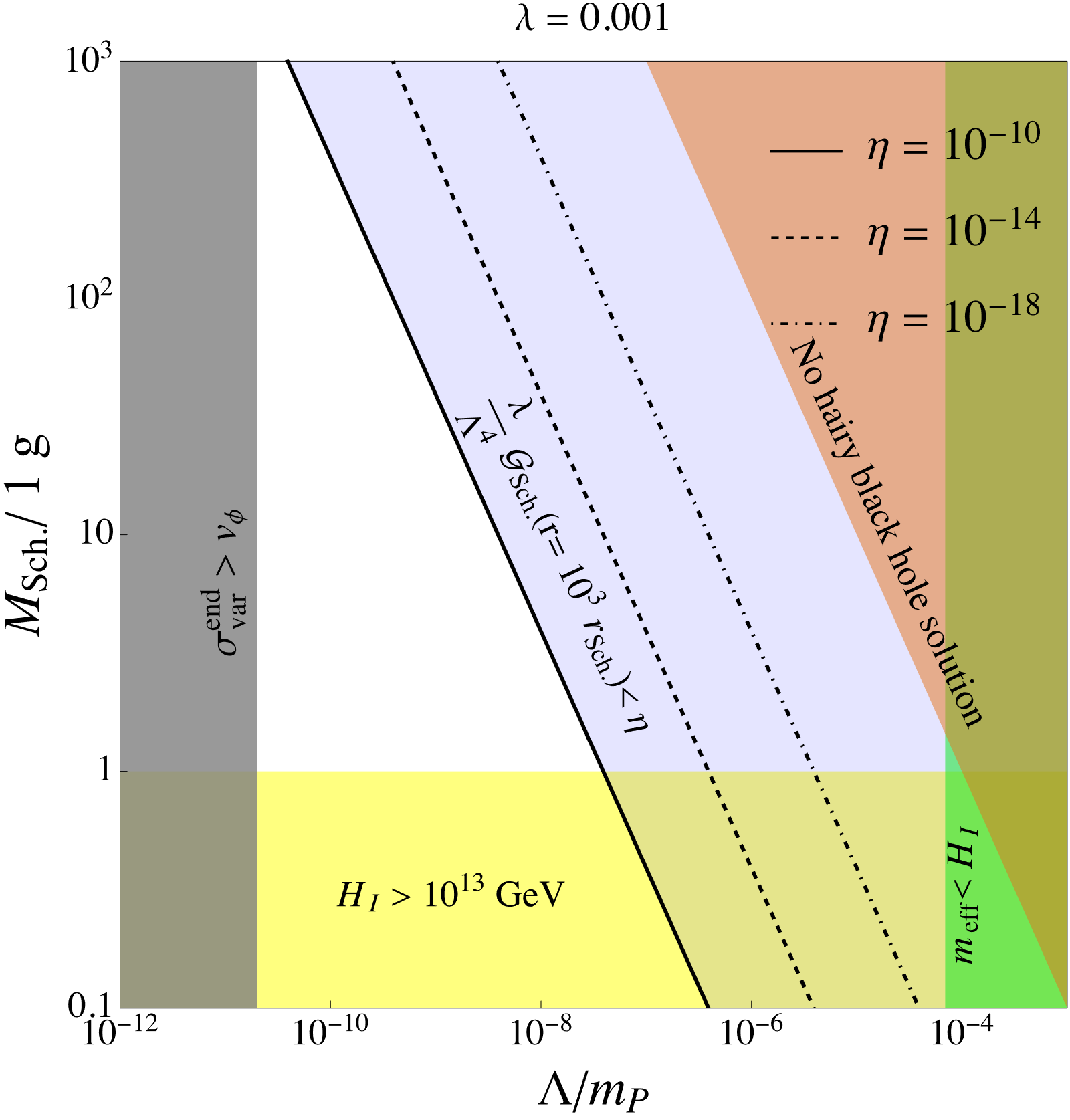}
\end{center}
\caption{Working parameter space (white region) for consistent hairy black hole solutions, in $\Lambda -M_{\rm Sch.}$ plane, for $\lambda=0.1$ (left panel) and $\lambda=0.001$ (right panel), considering $\alpha =1$. In the green-shaded region, the effective mass is lighter than the inflationary Hubble scale (cf. Eq. \eqref{eq:iso1}); while in the orange-shaded region, hairy black hole solutions don't exist (cf. Eq. \eqref{eq:l2bnd}). In the blue shaded region, the quartic  $\eta$ term would be more dominant near the black hole horizon (cf. Eq. \eqref{eq:L1bndBH}), beyond a distance of $r= 10^3 ~r_h$, for $\eta=10^{-10}$ (solid contour), with the constraint being relaxed for smaller values of $\eta$: $10^{-14}$ (dashed line)  and $10^{-18}$ (dot-dashed line). In the grey-shaded region, the variance of the field is larger than the background vev during inflation (cf. Eq. \eqref{eq:sigmavbnd}).} 
\label{fig:bndplt}  
\end{figure}

At this point, before delving into the tachyonic phase after inflation, we summarise the constraints we have obtained so far, which, as we will see, already tightly constrain our working parameter space for consistent hairy black holes. Fig. \ref{fig:bndplt} shows our working parameter space in the $M_{\rm Sch.}-\Lambda$ plane for $\lambda=0.1$ (left panel) and $\lambda=0.001$ (right panel). The shaded regions are excluded for the following reasons:
\begin{enumerate}
    \item Green-shaded region: This region violates the condition in Eq.~(\ref{eq:iso1}) that the scalar field lies in a stable minimum, with an effective mass satisfying $m_{\rm eff}\gtrsim H_I$ during inflation. In this region, the field is not sufficiently heavy and consequently, potentially significant isocurvature perturbations may survive. 
    \item Orange-shaded region: This region violates the instability condition given in Eq.~(\ref{eq:hbhcondtn}). When the effective GB coupling $\frac{\alpha}{\Lambda_2^2}$ lies below its critical value, Schwarzschild black hole would remain stable against scalar perturbations. Therefore, the instability required to trigger the phase transition from a Schwarzschild black hole to a hairy configuration does not arise.
    \item Blue-shaded region: This region violates the condition in Eq.~(\ref{eq:L1bndBH}), under which the bare scalar self-interaction is assumed to be negligible compared with the GB interaction near the black hole horizon. In this region, the self-interaction can no longer be consistently neglected.  Here, we choose $c_B=10^3$. These constraints are relaxed with a lower value of $\eta$, shown by dashed and dot-dashed contours.
    \item Grey-shaded region: This region lies outside the regime of validity of the perturbative treatment. For linear perturbation theory to remain applicable, the scalar-field fluctuations must be smaller than the background vacuum expectation value, such that the field remains localised at the vev during inflation. Equivalently, their variance must satisfy the condition derived in Eq.~(\ref{eq:sigmavbnd}).  
\end{enumerate}
Consequently, astrophysical hairy black holes requiring $\Lambda\sim10^{-19}\,\mathrm{GeV}$ (cf. Eq.~\eqref{eq:l2bmrk}) are already severely constrained by the requirement that the ESGB theory remain consistent with cosmological evolution. As we discuss next, the allowed parameter space becomes even more restricted once the post-inflationary tachyonic phase is taken into account.

% \cmb{(*}In the green-shaded region on the right, since the field is lighter than the Hubble rate during inflation, the field is not located at the minimum and there can be significant contribution to isocurvature perturbations (cf. Eq. \eqref{eq:iso1}). Above the red-shaded region, hairy BH solutions do not exist (cf. Eq. \eqref{eq:hbhcondtn}). In the blue-shaded region, the quartic $\eta$ term would be more dominant than the GB term close to the BH horizon beyond a certain distance, chosen here as $r=10^3 ~r_h$, which questions the validity of the hairy BH solutions. These constraints are relaxed with a lower value of $\eta$, shown by dashed and dot-dashed contours. Finally, in the grey-shaded region, the variance of the field becomes larger than the background vev. Thus, astrophysical hairy BHs with $\Lambda\sim 10^{-19}$ GeV (cf. Eq. \eqref{eq:lmbdaastro}) are severely constrained if we want the ESGB theory to be consistent with the cosmological history. These constraints are further restricted when the tachyonic phase after inflation is considered, which we discuss next. \cmb{ replaced. *)}

\textbf{Solution after inflation}:  We numerically solve the equation of motion for the mode function $f_{\kappa} (z)$ from the end of inflation ($z=0$). Since $M^2_{\rm eff}<0$ after inflation, small momentum modes that satisfy
\begin{align}
    \kappa^2 < |M^2_{\rm eff}(z)| 
\end{align}
can undergo tachyonic amplification. Thus, there is a time-dependent instability band in momentum space, which is given by
\begin{align}
   0< \kappa<\kappa_{\rm max} (z)\,,   \,\,\,\, \kappa_{\rm max} (z) \simeq \sqrt{ \frac{2 \alpha |\mathcal{G}_{\rm FLRW}(z)|a(z)^2}{\Lambda^2H_{\rm I}^2} }\,,\label{eq:tcbnd}
\end{align}
where modes with $\kappa> \kappa_{\rm max} (0)$ never undergo amplification. Modes with
\begin{align}
    \kappa \lesssim \kappa_{\rm hor} \simeq \frac{a (z) H(z)}{a_* H_{\rm I}}
\end{align}
are super-horizon modes, which grow outside the horizon.

\begin{figure}[h]
    \includegraphics[scale=0.4]{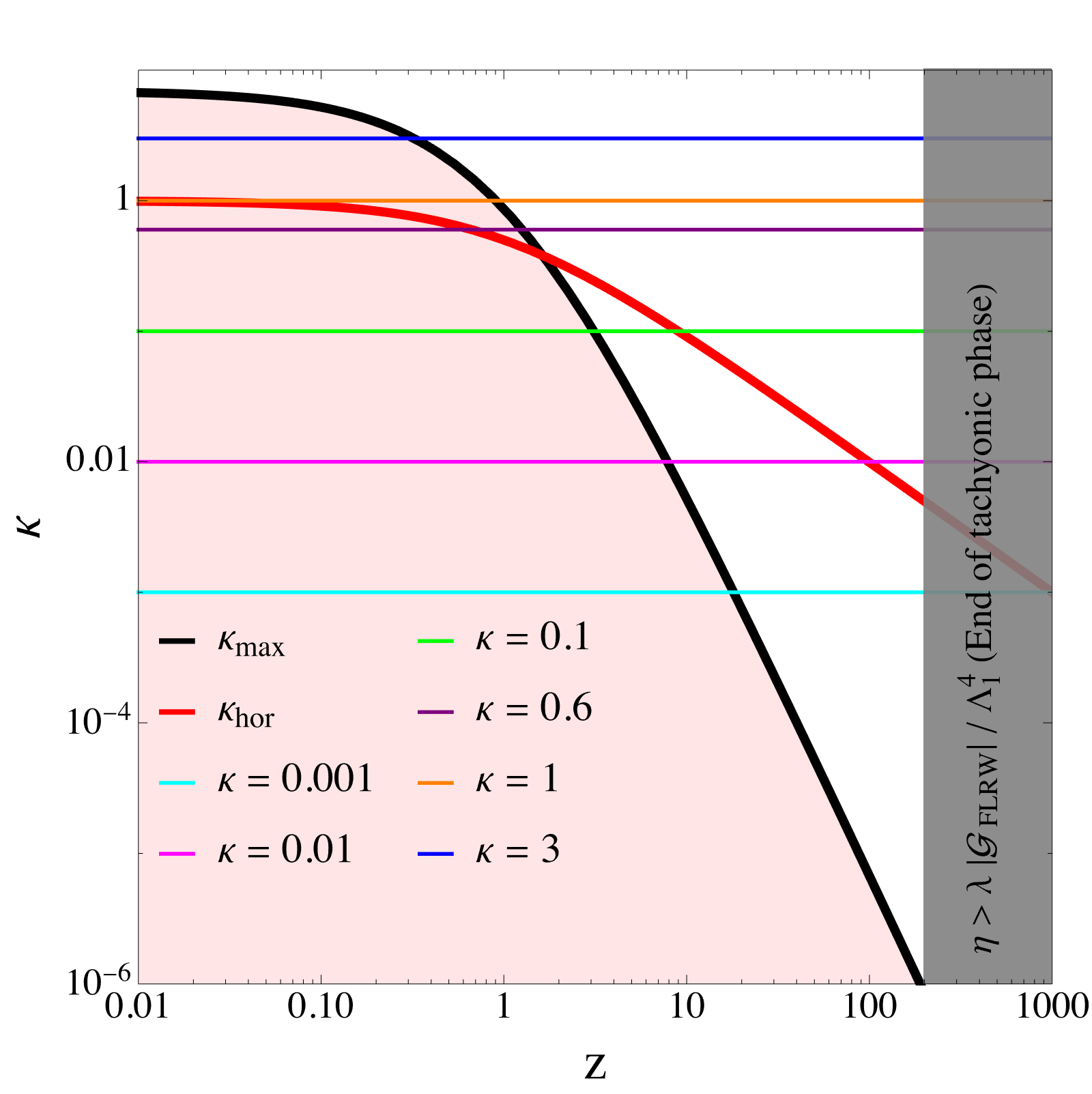} ~~
    \includegraphics[scale=0.4]{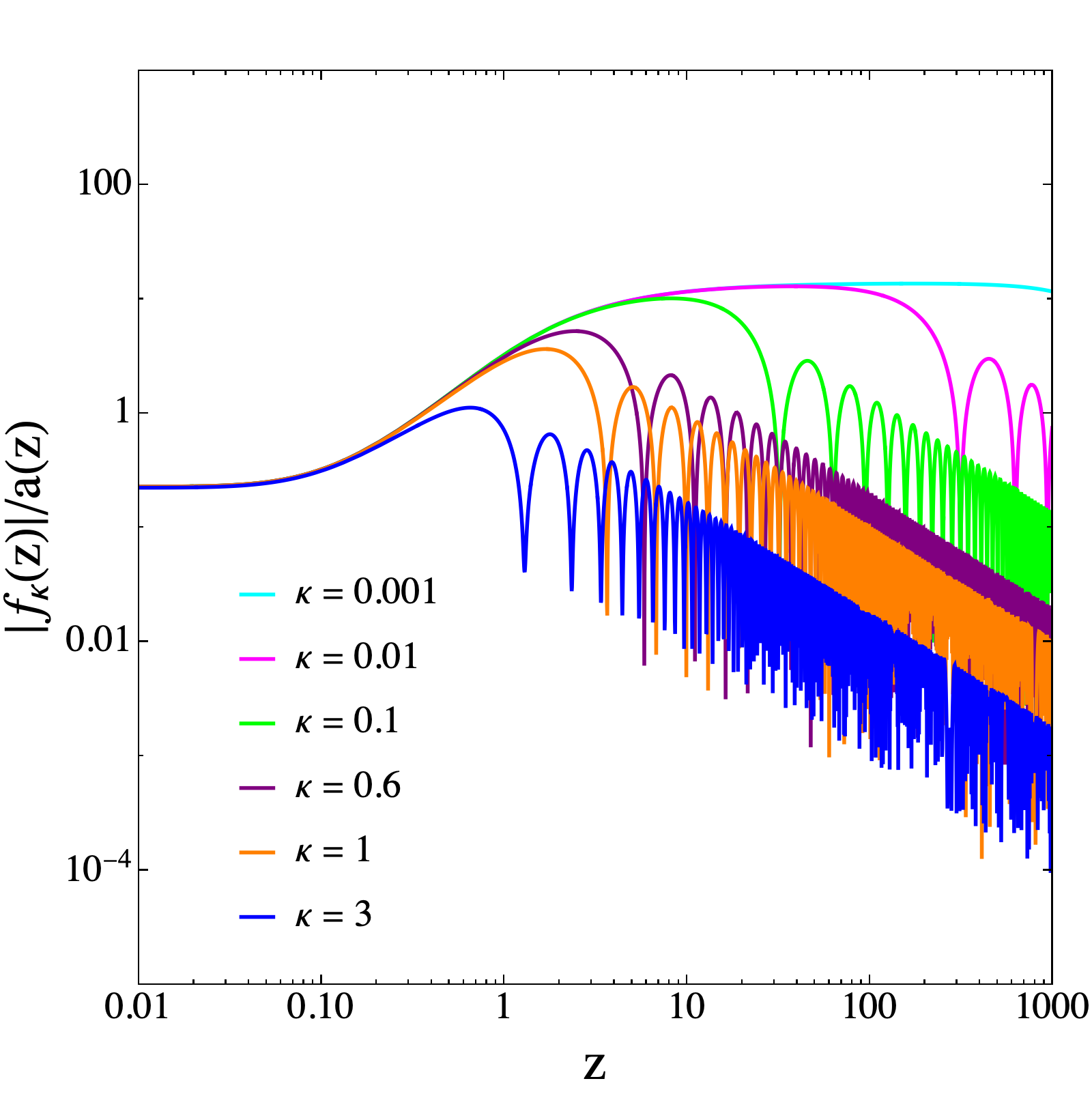}
\caption{\textit{Left panel:} The evolution of the tachyonic instability band (Eq. \eqref{eq:tcbnd}), shown by pink shaded region. The black and red thick contours indicate $\kappa_{\rm max}$ and $\kappa_{\rm hor}$, while the other contours correspond to different values of momentum modes, exiting the band at different epochs. The grey-shaded region indicates end of tachyonic phase for all modes. \textit{Right panel:} Evolution of $|f_{\kappa}|/a$ for the same values of $\kappa$ as in the left panel. We choose the benchmark: $\lambda=0.001, ~\alpha =1, ~ \Lambda=10^{-5}~m_P$.}  
\label{fig:fk}  
\end{figure}

Choosing the benchmark $\lambda=0.001, \alpha =1, \Lambda=10^{-5}~m_P$, we show the evolution of the tachyonic instability band in  the left panel of Fig. \ref{fig:fk}, indicated by the region shaded in pink. The thick black contour represents $\kappa_{\rm max}$, while the other colored contours represent different values of the momentum modes $\kappa$, exiting the instability band at different epochs, as $\kappa_{\rm max}$ decreases. The thick red contour corresponds to the comoving horizon size $\kappa_{\rm hor}$. The tachyonic phase ends in the grey-shaded region for all modes, as the quartic $\eta$ term becomes more dominant than the GB $\lambda$ term. The right panel of Fig. \ref{fig:fk} shows the evolution of $|f_{\kappa}|/a$, where the growth of the momentum modes present inside the instability band can be clearly seen.

\begin{figure}[t]
  \includegraphics[scale=0.4]{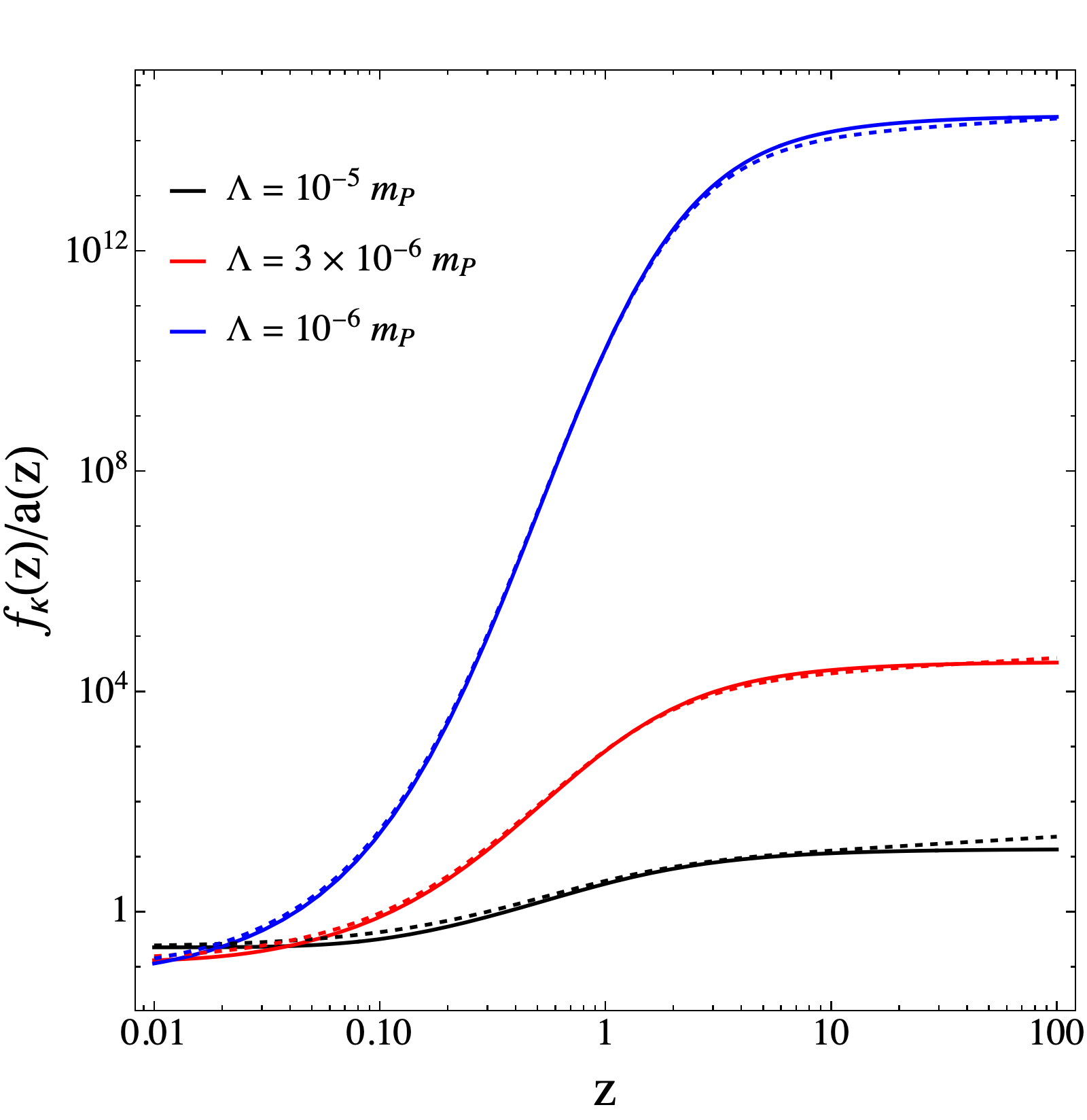}  ~~  \includegraphics[scale=0.6]{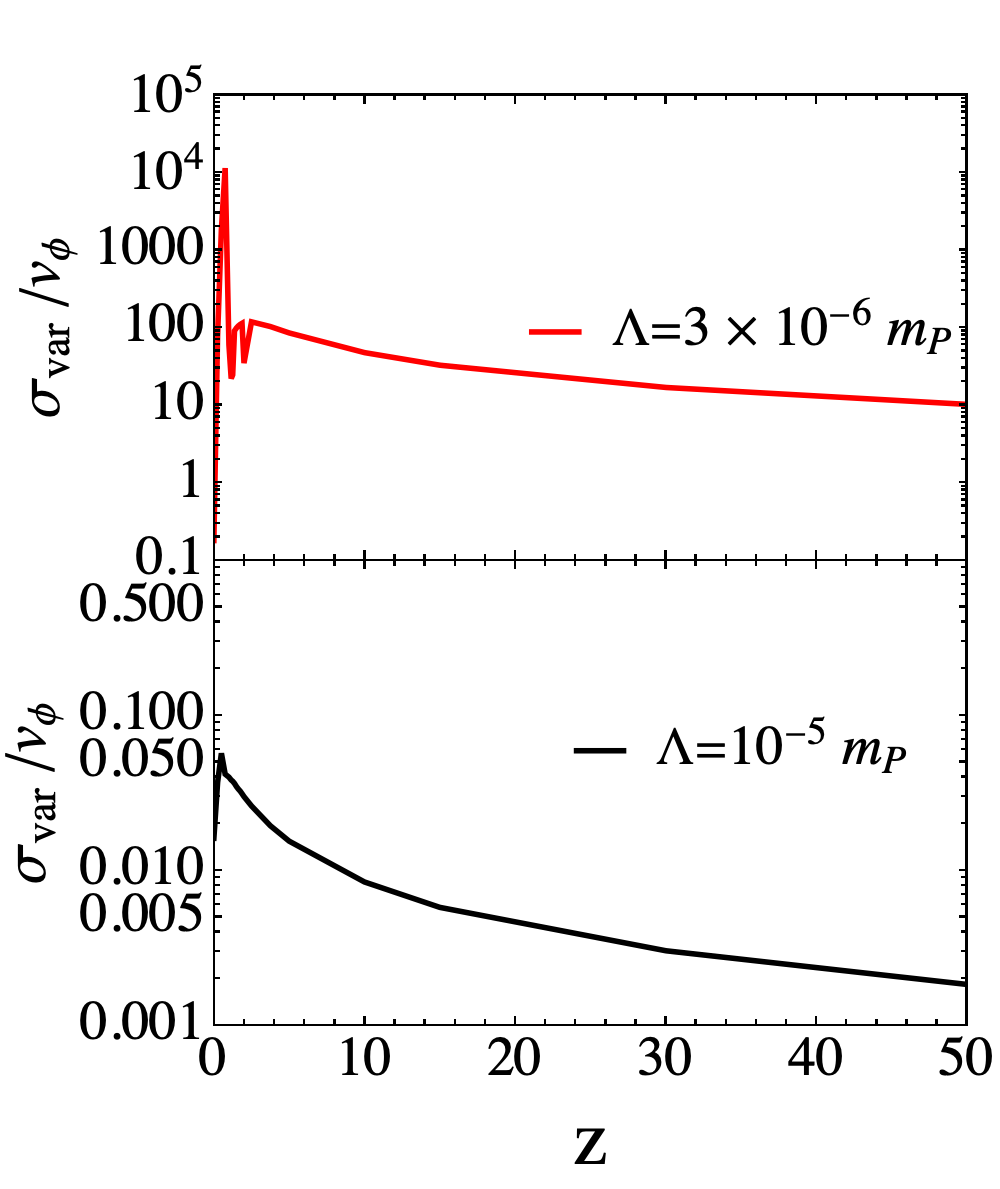}
\caption{\textit{Left panel:} Evolution of $|f_{\kappa}|/a$, for different values of $\Lambda$, considering $\kappa^2 \ll |M^2_{\rm eff} (z)|$. The dashed lines show the analytical estimates given by Eq. \eqref{eq:fkana}. \textit{Right panel:}  Evolution of $\sigma_{\rm var}$ (cf. Eq. \eqref{eq:sigmatach}) over the vev of the field $\phi$. We choose $\lambda= 0.001$, $\alpha=1$.} 
\label{fig:fk2}  
\end{figure}

Now, for $\kappa^2 \ll |M^2_{\rm eff} (z)|$, the mode functions can be estimated as
\begin{align}
    f_{\kappa} (z) &\sim f_{\kappa} (z_{\rm ini}) \mathcal{A} (z)\exp\left(\int_{z_{\rm ini}}^z \sqrt{|M^2_{\rm eff} (z)|} dz \right)\nonumber\\
    &\sim f_{\kappa} (z_{\rm ini}) \mathcal{A} (z) \exp\left( \int_{z_{\rm ini}}^z \frac{ 4\sqrt{3\alpha}H_I}{\Lambda(1+z)^3}  dz\right) \nonumber\\
    &\sim f_{\kappa} (z_{\rm ini}) \mathcal{A} (z)\exp\left(\frac{2\sqrt{3\alpha}H_I}{\Lambda}\left(1-\frac{1}{(1+z)^2}\right)\right)\,. \label{eq:fkana}
\end{align}
The WKB prefactor $\mathcal{A}(z)$ is a slowly growing function of $z$, compared to the exponential factor that is rapidly growing. The WKB approximation: $\left|\frac{M'_{\rm eff}(z)}{M_{\rm eff}^2 (z)}\right|\ll 1$, however, breaks down  as $z$ increases. For our purposes, we consider the ansatz $\mathcal{A}(z)= a(z)^{5/2}$, which fits considerably well with the full numerical solution in our regime of interest until the growth saturates. The left panel of Fig. \ref{fig:fk2} shows the evolution of $|f_{\kappa}|/a$, considering $\kappa^2 \ll |M^2_{\rm eff} (z)|$,  for different values of $\Lambda$.  As we can see, the growth is highly sensitive to the scale $\Lambda$. This is because the  maximum growth, estimated as
\begin{align}
    f_{\kappa}^{\rm Max }  \propto \mathcal{A}(z)\exp\left( \frac{2 \sqrt{3 \alpha}H_{\rm I}}{\Lambda}\right)\,,
\end{align}
gives, for instance,
\begin{align}
   \frac{\left|f_{\kappa}^{\rm Max }|_{\Lambda = 10^{-6} m_P}\right|}{\left|f_{\kappa}^{\rm Max }|_{\Lambda = 10^{-5} m_P}\right|} \sim  10^{19}\,,
\end{align}
considering $\alpha=1$. Thus, even a decrease in $\Lambda$ by less than $\mathcal{O} (1)$ causes a huge exponential growth of momentum modes inside the tachyonic band. 

As a result of this exponential tachyonic growth, the linear approximation breaks down very soon. Precisely, this would happen when the first leading non-linear term  in our  expansion of $V'(\phi)$  in the equation of motion (cf. Eq. \eqref{eq:eompert}) becomes dominant, i.e. $\frac12 V'''(\phi) (\delta \phi)^2 \gtrsim V''(\phi) \delta \phi$. Considering $\phi=v_{\phi }$, this gives the condition that $\sigma_{\rm var}(z) \lesssim v_{\phi}$, for our linear analysis to remain valid. The right panel of Fig. \ref{fig:fk2} shows  $\sigma_{\rm var}$ for two values of $\Lambda$, where 
% \begin{align}
%     <(\delta \phi(z)-\delta \phi (z_{\rm ini}))^2>
% \end{align}
\begin{align}
   \sigma_{\rm var} (z) =\sqrt{<\delta \hat{\phi}^2 (z)>}\simeq \sqrt{H_I^2\int_0^{\kappa_{\rm max}} \frac{d^3\kappa}{(2 \pi)^3} \frac{|f_\kappa(z)|^2}{a^2(z)}}\,, \label{eq:sigmatach}
\end{align}
found by integrating numerically over all the relevant tachyonic momentum modes. $\sigma_{\rm var} (z)$ increases initially due to the tachyonic growth, but decreases later due to redshift.  The tachyonic growth for $\Lambda=10^{-5}$ is not so efficient, such that $\sigma_{\rm var}$  remains less than $v_{\phi}$. While for $\Lambda= 3 \times 10^{-6} ~m_P$, since the growth is large, the linear approximation breaks down. At this point, our analysis becomes invalid and a more detailed non-linear simulation becomes essential. The field is expected to enter a runaway regime carrying large energy density, with the potential effectively being unbounded with a subdominant stabilising $\eta$ term. The energy density of the field, obtained from the energy-momentum tensor $T^{\mu}_{~\nu}$, is given as
\begin{align}
    \rho_{\phi}= T_{00}= \frac{\dot{\phi}^2}{2} + \frac{(\nabla \phi)^2}{2 a^2} + \frac{\eta}{4} \phi^4 -8 H^2\left(\frac{\nabla^2 f}{a^2} - 3 H \frac{\partial f}{\partial t}\right)\,, \label{Eq:rhophi}
\end{align}
where $f= \frac{\lambda}{4 \Lambda_1^4} \phi^4 -\frac{\alpha}{2 \Lambda_2^2} \phi^2$ is the function coupled to the GB term. The first three terms in Eq. \eqref{Eq:rhophi} are the kinetic, gradient and potential contributions respectively, while the last term represents the contribution from interaction associated with the curvature-induced GB coupling. As found in analogous studies using lattice simulations \cite{Laverda:2026slq}, the contribution from the interaction term is expected to rapidly increase during the tachyonic stage, until $\mathcal{G}_{\rm FLRW}(z)$ redshifts enough. The energy  density in the inhomogeneities is then converted into the kinetic and gradient terms. Finally, the potential contribution would become important when the symmetry is restored. However, with a smaller value of $\Lambda$ leading to a large tachyonic growth, the energy density is expected to exceed the background energy density of the Universe much before the potential stabilizes.
\begin{figure}[h]
\begin{center}
  \includegraphics[scale=0.4]{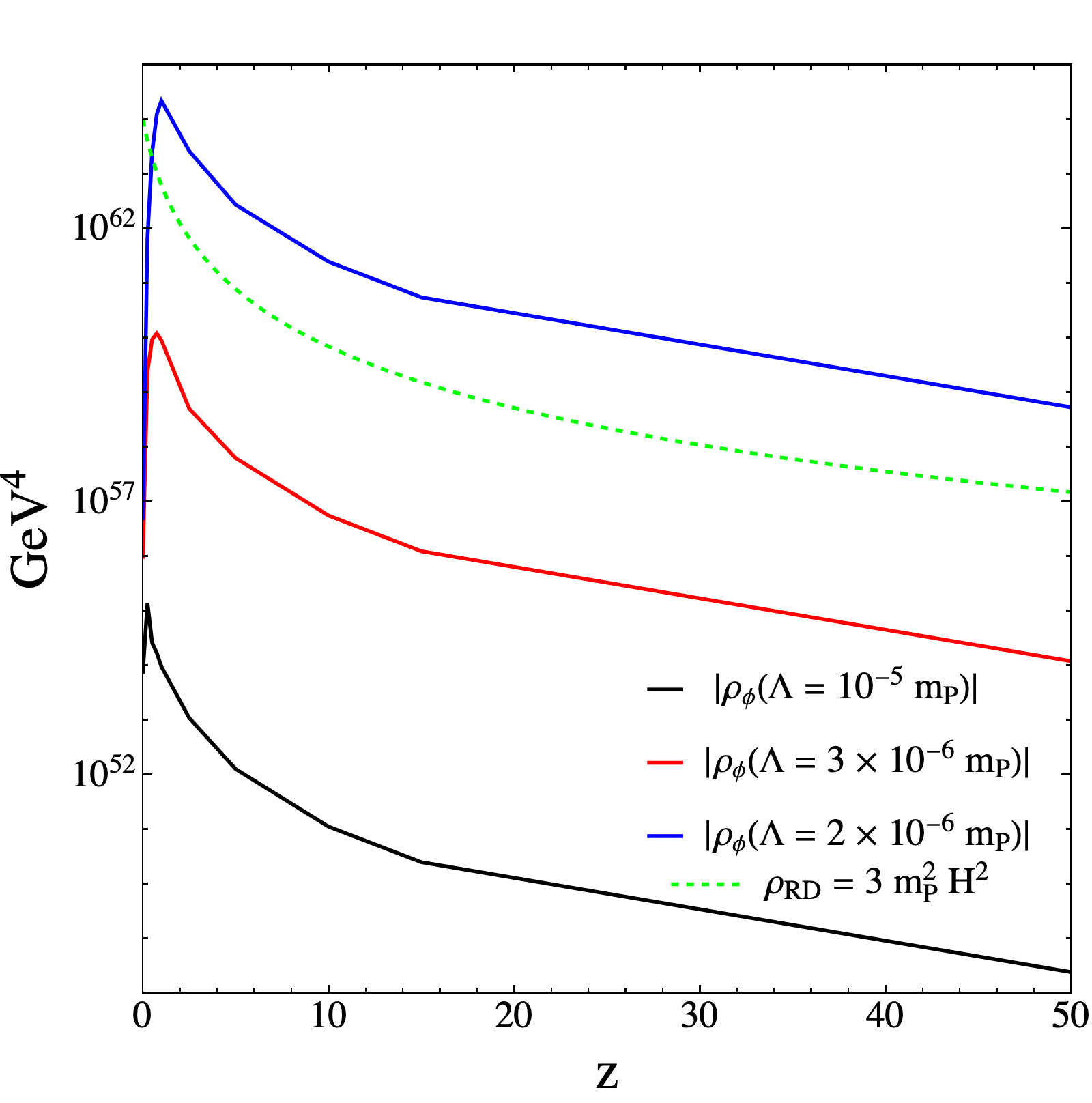}  
\caption{Evolution of the inhomogeneous energy density of the field $\phi$ (cf. Eq. \eqref{eq:rhophifl}), along with that of the background energy density of the Universe shown by the green dashed line. We choose $\lambda=0.001,~ \alpha=1$.} 
\label{fig:rhophi}  
\end{center}
\end{figure} 

From Eq. \eqref{Eq:rhophi}, the inhomogeneous energy density in the fluctuations $\delta \phi$ is calculated to be
\begin{align}
    \rho_{\phi} (z) \simeq &\left(\frac{H_I}{a(z)}\right)^4\int_{\kappa_{\rm min}}^{\kappa_{\rm max}} \frac{d^3 \kappa}{(2 \pi)^3}\Bigg[\frac12 |f'_{\kappa}(z)-\frac{a'(z)}{a(z)}f_{\kappa}(z)|^2 +\frac12 \kappa^2 |f_{\kappa}(z)|^2+\frac32 \eta ~Y(z)^2|f_{\kappa}(z)|^2 \Bigg] \nonumber\\
    & + 24 H(z)^3 \left(\frac{H_I}{a(z)}\right)^3 \int_{\kappa_{\rm min}}^{\kappa_{\rm max}} \frac{d^3 \kappa}{(2 \pi)^3} \Bigg[\frac{\lambda}{\Lambda_1^4}\left(\frac{H_I}{a(z)}\right)^3 \bigg(3 Y'(z)Y(z)|f_{\kappa}(z)|^2+3 Y(z)^2 ~\text{Re}[f^*_{\kappa}(z)f'_{\kappa}(z)]\nonumber \\ &- 6Y(z)^2 |f_{\kappa} (z)|^2\frac{a'(z)}{a(z)}\bigg)  
     -\frac{\alpha}{\Lambda_2^2}\left(\text{Re}[f^*_{\kappa}(z)f'_{\kappa}(z)]-\frac{a'(z)}{a(z)}|f_{\kappa} (z)|^2 \right) \Bigg]\,,\label{eq:rhophifl}
\end{align}
where we integrate over the sub-horizon tachyonic modes. The second term within the square brackets indicates the GB contribution. In Fig. \ref{fig:rhophi}, we show the evolution of the energy density of the field $\phi$ found using Eq. \eqref{eq:rhophifl}, for three values of $\Lambda$, along with the background energy density of the Universe (green dashed line) considering RD domination. Clearly, for lower values of $\Lambda$, the increase in the energy density is much larger. For $\Lambda = 2 \times 10^{-6}m_P$, the energy density even exceeds that of the RD background, which may be catastrophic for the cosmological history\footnote{To be precise, with a negligible bare mass term, the energy density in the fluctuations of $\phi$ scale relativistically, soon after the tachyonic growth, and hence may lead to a large contribution to the radiation components of the Universe, which is constrained from BBN observations.}.  As a conservative estimate, we consider our analysis to be valid until the linear approximation breaks down (cf. right panel of Fig. \ref{fig:fk2}), when our above estimation of  energy density, say for $\Lambda= 3\times 10^{-6} m_P$, is also expected to become invalid. A complete analysis would involve non-linear evolution of fluctuations and requires a detailed lattice simulation \cite{Laverda:2026slq}.  Hence, for the tachyonic growth to not disturb the standard cosmological history, we consider the cutoff scale to be close to $10^{-5} m_P$, which at the same time could also lead to hairy black hole solutions for ultralight PBH  with $M_{\rm  Sch.}\sim \mathcal{O}(10)~\text{g}$ (cf. Eq. \eqref{eq:l2bnd}). In such a case, the tachyonic growth is expected to be inefficient, and the field finally stabilises when the $\eta$ term dominates. We next discuss the possible fate of such ultralight PBHs in the above cosmological settings.

\section{Chronology of hairy PBH}\label{Sec5}
\begin{figure}[h]
\begin{center}
  \includegraphics[scale=0.4]{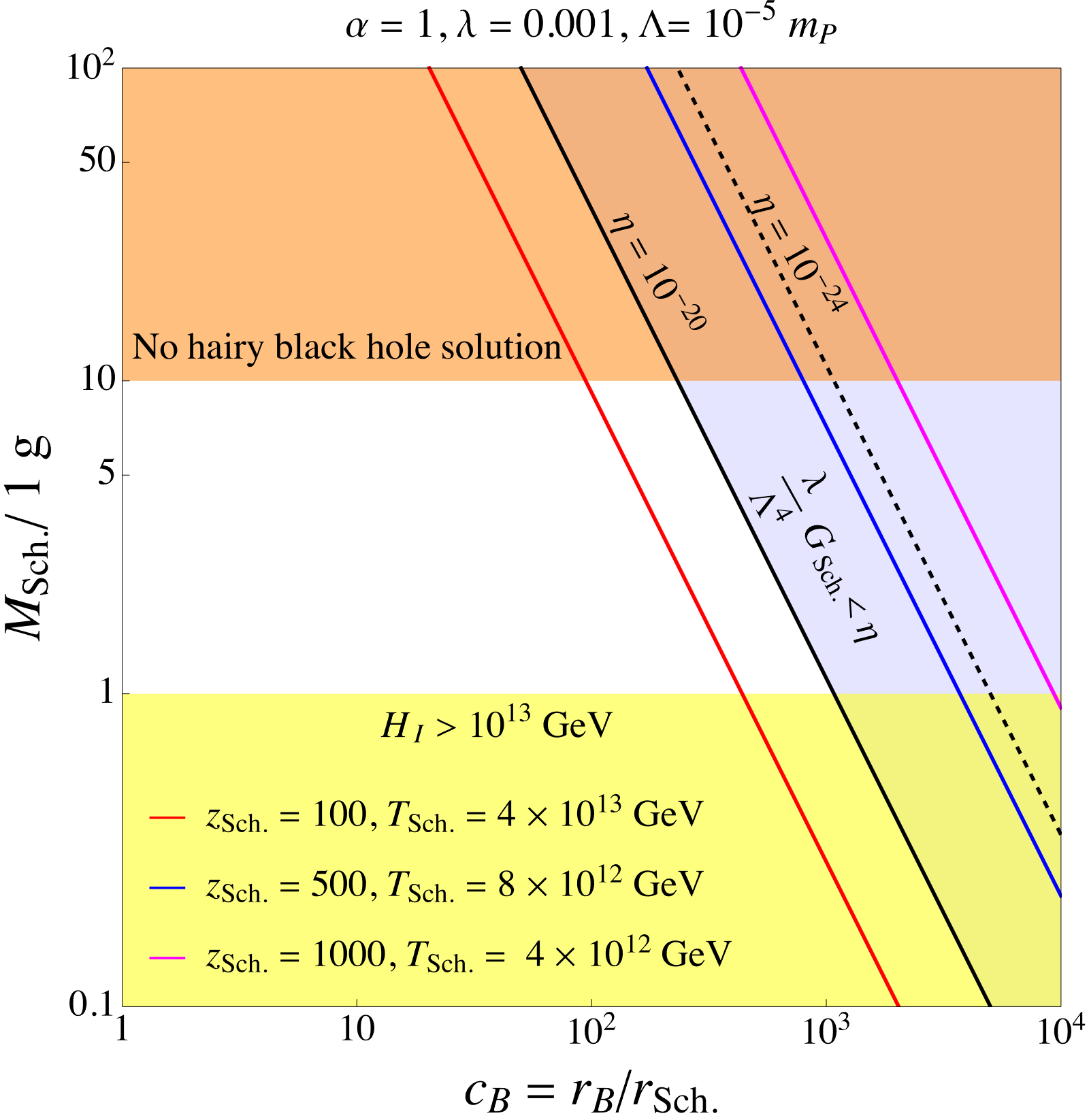}  
\caption{Allowed PBH mass range (white region), as a function of the dimensionless distance from the black hole horizon: $c_B=r_B/r_{\rm Sch.}$. The colored lines indicate the values of $z_{\rm Sch.}$, as estimated in Eq. \eqref{eq:zsch}. As the Universe expands (larger $z_{\rm Sch.}$), region that can be approximated by the Schwarzschild solution increases (larger $c_B$). The orange-shaded region doesn't lead to hairy black hole solutions (cf. Eq. \eqref{eq:L1bndBH}), while the yellow-shaded region violates the lower bound on PBH mass given by Eq. \eqref{Eq:PBHlowbnd}.} 
\label{fig:zsch}  
\end{center}
\end{figure} 
%In the previous section, we established that, in order to \cmb{suppress} efficient tachyonic growth, the BH mass range that can develop scalar hair should  be in the ultralight PBH mass range: $\mathcal{O} (10)$ g. We discuss the chronology of such PBHs here. The initial mass of PBH during formation\footnote{PBH can be formed in the early Universe through several mechanisms; see \cite{Carr:2020gox, Escriva:2021aeh} for a comprehensive review. For our purpose, the analysis is independent of the specific formation mechanism; hence we don't discuss PBH formation details here.} at a time $t_{\rm form}$ can be related to the mass enclosed in the particle horizon, as \cite{Carr:1974nx}

In the previous section, we established that, in order to suppress efficient tachyonic growth, black holes capable of developing scalar hair must lie in the ultralight PBH mass range, with $M_{\rm PBH}\sim\mathcal{O}(10)\,\mathrm{g}$. We now discuss the cosmological history of such PBHs. The initial mass of a PBH formed at time $t_{\rm form}$\footnote{PBHs can form in the early Universe through a variety of mechanisms; see Refs.~\cite{Carr:2020gox,Escriva:2021aeh} for comprehensive reviews. Since our analysis does not depend on the specific formation mechanism, we do not discuss the details of PBH formation here.} can be related to the mass enclosed within the particle horizon at that time as 

\begin{align}
    M_{\rm PBH}^{\rm form} = \frac43 \pi \gamma \frac{\rho_{\rm total} (t_{\rm form})}{H^3(t_{\rm form})} \,,
\end{align}
where $\rho_{\rm total}(t)= 3 m_P^2 H(t)^2$ is the total energy density of the universe, and  $\gamma$ is an $\mathcal{O}(1)$ parameter characterising the efficiency of gravitational collapse. During radiation domination, it is typically taken to be $\gamma\simeq0.2$ \cite{Carr:2009jm}. Using the upper bound on the Hubble scale at the end of inflation, i.e. $H_I \lesssim 10^{13}$ GeV \cite{Planck:2018jri}, we get a lower bound on the possible PBH mass formed during such a collapse as
\begin{align} 
 M_{\rm PBH}^{\rm form} \gtrsim  1 ~\text{g} \,,\label{Eq:PBHlowbnd}
\end{align}
where we used $\rho_{\rm total}(t_{\rm form})= 3 m_P^2 H_I^2$. Assuming that the Universe is RD after inflation, the total energy density is given by
$\rho_{\rm total}(t)=\rho_{\rm rad}(t)\equiv \frac{\pi^2}{30}g_*(T)T^4(t)$,
where $g_*(T)\simeq106.75$ denotes the effective number of relativistic degrees of freedom.
%Considering a RD Universe after inflation, $\rho_{\rm total} (t)=\rho_{\rm rad} (t) \cmb{\equiv} \frac{\pi^2}{30}g_* T(t)^4$ denotes the radiation energy density, where $g_*\sim 100$ is the relativistic degrees of freedom. 
Thus, the temperature of the radiation bath at the time of PBH formation is found to be
\begin{align}
    T(t_{\rm form}) \simeq 1.7 \times 10^{-3} ~m_{P} \left(\frac{1 ~\text{g}}{ M_{\rm PBH}^{\rm form}}\right)^{1/2}\,.
\end{align}
The temperature can be related to our variable $z$ using
\begin{align}
    T(z) = \left(\frac{90}{\pi^2 g_*}\right)^{1/4} \frac{\sqrt{m_P H_I}}{1+z}\,.
\end{align}
Thus, we have 
\begin{align}
    z(T_{\rm form})= z_{\rm form}\simeq   3.2 \times 10^{2} \left(\frac{ M_{\rm PBH}^{\rm form}}{1~\text{g}}\right)^{1/2}\left(\frac{H_{\rm I}}{m_{\rm P}}\right)^{1/2}-1\,.
\end{align}
For $M_{\rm PBH}^{\rm form}=10~ \text{g}$, we get $z_f \simeq 1.05$

%Now, at the time of formation, the choice of metric that adequately describes PBH is debatable and may be non-trivial \cite{Faraoni:2018xwo, Harada:2021xze, Picker:2021jxl, Hutsi:2021nvs, Boehm:2021kzq}. This is because during this formation epoch in the early Universe, the Hubble horizon is still smaller and comparable to the PBH horizon radius. However, as the Universe expands and the Hubble horizon becomes sufficienly large, one can safely assume the Schwarzschild metric solution close to the BH. For our purpose of investigating the formation of hairy black holes via SSB, we consider that the initial PBH metric can be estimated by the Schwarzchild solution, until a distance of $r\sim 10^{2-3}~ r_{\rm Sch.}$ away from the BH horizon. Let us estimate the epoch from which this assumption holds.    

At the time of PBH formation, the appropriate spacetime geometry is subtle and remains a subject of debate \cite{Faraoni:2018xwo,Harada:2021xze,Picker:2021jxl,Hutsi:2021nvs,Boehm:2021kzq}. Indeed, during this early epoch, the PBH horizon radius constitutes a non-negligible fraction of the Hubble radius, so a clear separation between the local black hole geometry and the cosmological background is not yet available. As the Universe expands, however, the Hubble radius becomes much larger than the length scales relevant to the black hole, and the near-PBH geometry can be safely approximated by the Schwarzschild solution. For the purpose of studying the formation of hairy black holes through spontaneous symmetry breaking, we assume that the spacetime surrounding the PBH is well described by the Schwarzschild metric out to a radial distance of $r\sim 10^{2\text{--}3}r_{\rm Sch.}$ where $r_{\rm Sch.}$ denotes the Schwarzschild radius. At larger distances, i.e. $r\gtrsim10^{2\text{--}3}r_{\rm Sch.}$, the spacetime is taken to approach the cosmological FLRW geometry. We now estimate the epoch from which this approximation becomes valid.

%Firstly, in order for the Schwarzchild solution to be valid until $r_B=c_B ~r_{\rm Sch.}$ away from the black hole, we expect the effect of the GB contribution arising because of the BH spacetime to be more dominant compared to the contribution from that of the FLRW spacetime, i.e. $\mathcal{G}_{\rm Sch.}(r) \gtrsim \mathcal{G}_{\rm FLRW}(z)$. This happens at an epoch, which is estimated as

If we define the outer boundary of the region within which the Schwarzschild approximation remains valid as $r_B\equiv c_B r_{\rm Sch.}$, the GB contribution from the black hole at $r=r_B$ should dominate over that of the FLRW background, i.e. $\mathcal{G}_{\rm Sch.}(r_B) \gtrsim \mathcal{G}_{\rm FLRW}(z)$. At the epoch characterised by $z$, this condition can be estimated as
\begin{align}
    z_{\rm Sch.} \gtrsim  3.3 \times 10^2 \left(\frac{H_I}{m_P}\right)^{1/2} \left(\frac{M_{\rm Sch.}}{1 ~\text{g}}\right)^{1/2} c_{B}^{3/4} -1\,. \label{eq:zsch} 
\end{align}
%\cmb{(*} In addition, as discussed earlier in Section \ref{sec:setup}, in order to describe our hairy BH solution, the $\lambda$ term has to be more dominant compared to the $\eta$ term, which is given by the condition in Eq. \eqref{eq:L1bndBH}. In Fig. \ref{fig:zsch} , we show the range of parameters, where our analysis is expected to hold. The $X$ axis indicates the distance away from the PBH in units of $r_{\rm Sch.}$, while the $Y$ axis denotes the mass of the Schwarzchild PBH. Considering $\Lambda = 10^{-5} m_P$ in order to avoid significant tachyonic growth,  $M_{\rm Sch.}\gtrsim \mathcal{O} (10)$ g (orange shaded region) would not lead to hairy BH solution according to the condition given in Eq. \eqref{eq:hbhcondtn}. \cmb{*) (I think we dont need to repeat this argument here.)} 
This condition implies that, if the outer boundary of the region described by the Schwarzschild geometry is defined as $r_B=c_Br_{\rm Sch.}$, the local Schwarzschild region can be consistently distinguished from the surrounding FLRW spacetime for $z\gtrsim z_{\rm Sch.}$. This requirement gives an estimate of the distance and time scales in the early Universe, in which our parameter space shown in Fig.~\ref{fig:bndplt} could work. Fixing $\Lambda=10^{-5}m_P$, which doesn't lead to efficient tachyonic growth as shown earlier, we investigate the allowed PBH mass range as a function of $c_B=r_B/r_{\rm Sch.}$. The resulting parameter space is shown in Fig.~\ref{fig:zsch}. The orange-shaded region violates the instability condition for a Schwarzschild black hole given in Eq.~\eqref{eq:hbhcondtn}, while the red-blue-magenta colored contours represent the values of $z_{\rm Sch.}=100, 500,$ and $1000$ respectively, determined by Eq.~\eqref{eq:zsch}. As the Universe evolves and $z$ increases, the Hubble radius grows, allowing the region well approximated by the Schwarzschild geometry to extend to progressively larger radii (towards right in the figure). Consequently, the effects of cosmological expansion become increasingly negligible in the vicinity of the PBH. The shaded region in blue indicates the dominance of the  $\eta$ term (cf. Eq. \eqref{eq:L1bndBH}), considering $\eta= 10^{-20}$ (solid line) and $10^{-24}$ (dashed line).

%Values of $r$ lying to the left of these contours can be approximated by the Schwarzchild solution. \cmb{(-- ?)} As time passes, the Hubble horizon increases and the effect of  the expansion of the Universe is negligible close to the PBH. Due to the decrease of $\mathcal{G}_{\rm{FLRW}}$ with time, distances further away from the PBH can now be approximated by the Schwarzchild solution. Thus, the $z_{\rm{Sch.}}$ contours shift to the right for larger $z_{\rm{Sch.}}$. The shaded region in green indicates the dominance of the  $\eta$ term, which can be relaxed with a lower value of $\eta$ shown by the dashed line. \cmb{(-- ?)}

%For our parameter range of interest, we find that the symmetry of the \sout{potential} \cmb{vacuum} is restored in the FLRW background after $z>z_{\rm Sch.}$ \cmb{since the self-interaction term becomes dominant as depicted in Fig. \ref{fig:potFLRW}-(c)}. The field then  oscillates around $\phi=0$ \cmb{via equations of motion and this behaviour is plotted in Fig. \ref{fig:hmo}.} \sout{Thus}, \cmb{On the other hand,} Schwarzchild BHs with $\phi\big|_{r=r_{\rm Sch.} }=0$, would become unstable and are expected to go through a phase transition to hairy BHs with $\phi\big|_{r=r_{\rm Sch.} }=v_{\phi}$ \cite{Latosh:2023cxm}. 

For our parameter range of interest, we find that the symmetry is restored in the FLRW background after $z>z_{\rm Sch.}$, as the scalar self-interaction becomes dominant, as illustrated in Fig.~\ref{fig:potFLRW}-(c). The field subsequently oscillates about the symmetric vacuum at $\phi=0$, as demonstrated by the numerical solution of the equation of motion shown in Fig.~\ref{fig:hmo}. In contrast, Schwarzschild black holes with $\phi(r=r_{\rm Sch.})=0$ become unstable and are expected to undergo a phase transition to hairy configurations characterised by a nonzero horizon value, $\phi(r=r_{\rm Sch.})=v_\phi$ \cite{Latosh:2023cxm}.

%The time-scale of this phase transition can be estimated by the analysis of the quasi-normal modes, which was studied in detail in \cite{Hyun:2024sfv}. The perturbations of the scalar field around the BH is given by $e^{-i \omega t}$, where $\omega$ is complex, taking discrete values. Thus, modes with Im[$\omega] >0$, would lead to a growing solution. In \cite{Hyun:2024sfv}, such a mode was found to appear for a Schwarzschild BH, when $\text{Im}[G M_{\rm Sch.}\omega_n]\sim 10^{-5}$. Considering the growth of the perturbations to be dominant when $\omega t\sim \mathcal{O}(1)$, the time scale of the phase transition from Schwarzchild BH to hairy BH is thus given by

The characteristic timescale of this phase transition can be estimated from the quasinormal mode analysis performed in~\cite{Hyun:2024sfv}. Scalar perturbations around the black hole behave as $e^{-i\omega_n t}$, where the complex frequencies $\omega_n$ form a discrete spectrum. With this convention, a mode satisfying $\operatorname{Im}(\omega_n)>0$ grows exponentially and signals an instability of the Schwarzschild solution. Ref.~\cite{Hyun:2024sfv} identified such an unstable mode with a dimensionless growth rate of $\operatorname{Im}(G M_{\rm Sch.}\omega_n)\sim10^{-5}$. Taking the perturbation to become significant when $\operatorname{Im}(\omega_n)t\sim\mathcal{O}(1)$, the corresponding instability-growth timescale, which we use as an estimate of the transition time from a Schwarzschild black hole to a hairy black hole, is given by

\begin{align}
    t_h -t_0 \sim 4 \times 10^{-10} \left(\frac{M_{\rm Sch.}}{1 ~\text{g}}\right) \text{GeV}^{-1}\,,\label{eq:thair}
 \end{align}
where $t_0$ indicates the time when $\phi(r_{\rm Sch.})=0$ and can be considered around the epoch when the field starts to oscillate in the FLRW background. Using $dt=(1+z)~ dz/ H_I$, in terms of $z$, we get 
\begin{align}
    z_h \sim (1+z_0)\sqrt{1+2H(z_0)(t_h-t_0)} -1\,,\label{eq:zhair}
\end{align}
where $H(z_0)= \frac{H_I}{(1+z_0)^2}$. Now, $z_0$ can be estimated to be around the epoch when $H(z_0)$ becomes equal to the saturated value of  $m_{\rm eff} (z_0)\simeq \sqrt{\frac{6 \eta \alpha}{\lambda}} \Lambda$ (cf. Fig. \ref{fig:hmo}). Thus,
\begin{align}
    z_0 \sim  1.14 \times 10^{4}\left(\frac{H_I}{\Lambda}\right)^{1/2} \left(\frac{\lambda}{0.001}\right)^{1/4} \left(\frac{1}{\alpha}\right)^{1/4} \left(\frac{10^{-20}}{\eta}\right)^{1/4}\,.\label{eq:z0}
\end{align}
From Eqs. \eqref{eq:thair}, \eqref{eq:zhair}, \eqref{eq:z0}, we thus see that $z_h \sim z_0$ for our choice of parameters. Hence, the Schwarzschild black hole is expected to transform almost instantly to the hairy black hole in the symmetry-broken phase, once it is unstable at $\phi (r_{\rm Sch.})=0$, much before it evaporates\footnote{Here, we consider the standard evaporation temperature of a Schwarzschild black hole for this comparison, which is given as $T_{\rm ev} \sim \frac{m_P^{5/2}}{M_{\rm Sch.}^{3/2}}$ \cite{Carr:2009jm, Carr:2020gox}. This gives $T_{\rm ev}\sim 4 \times10^{8}~\text{GeV}$ for $M_{\rm Sch.}= 10~\text{g}$, which corresponds to $z (T_{\rm ev})\sim 6 \times 10^{6}\gg z_h$, for our parameters $\Lambda= 10^{-5} m_P,~ \alpha=1,~ \lambda=0.001$. However, it should be noted that for hairy black holes, the evaporation temperature might get modified, although we expect the change to be small enough.}. 

\begin{figure}
    \includegraphics[scale=0.4]{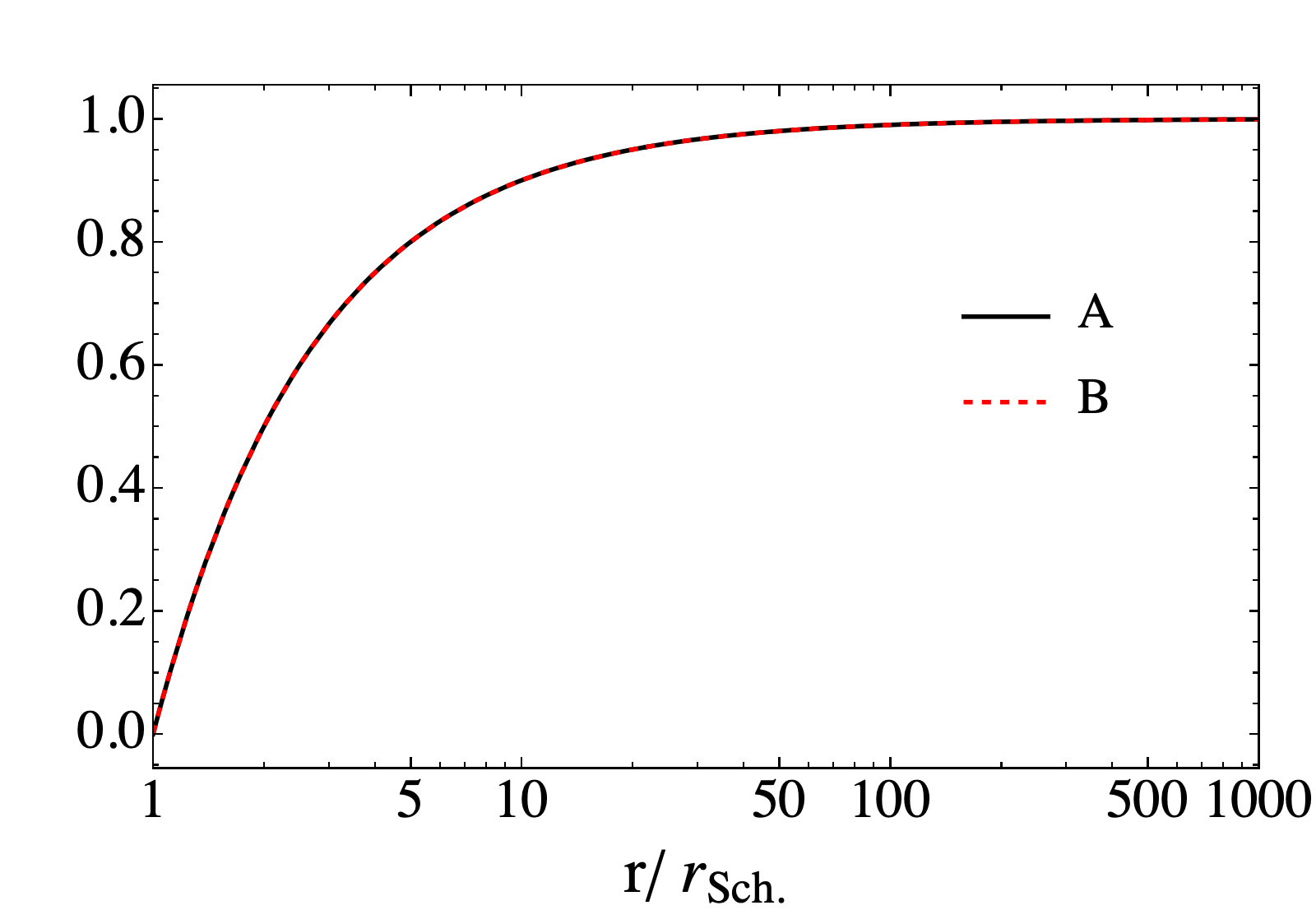} ~~
    \includegraphics[scale=0.4]{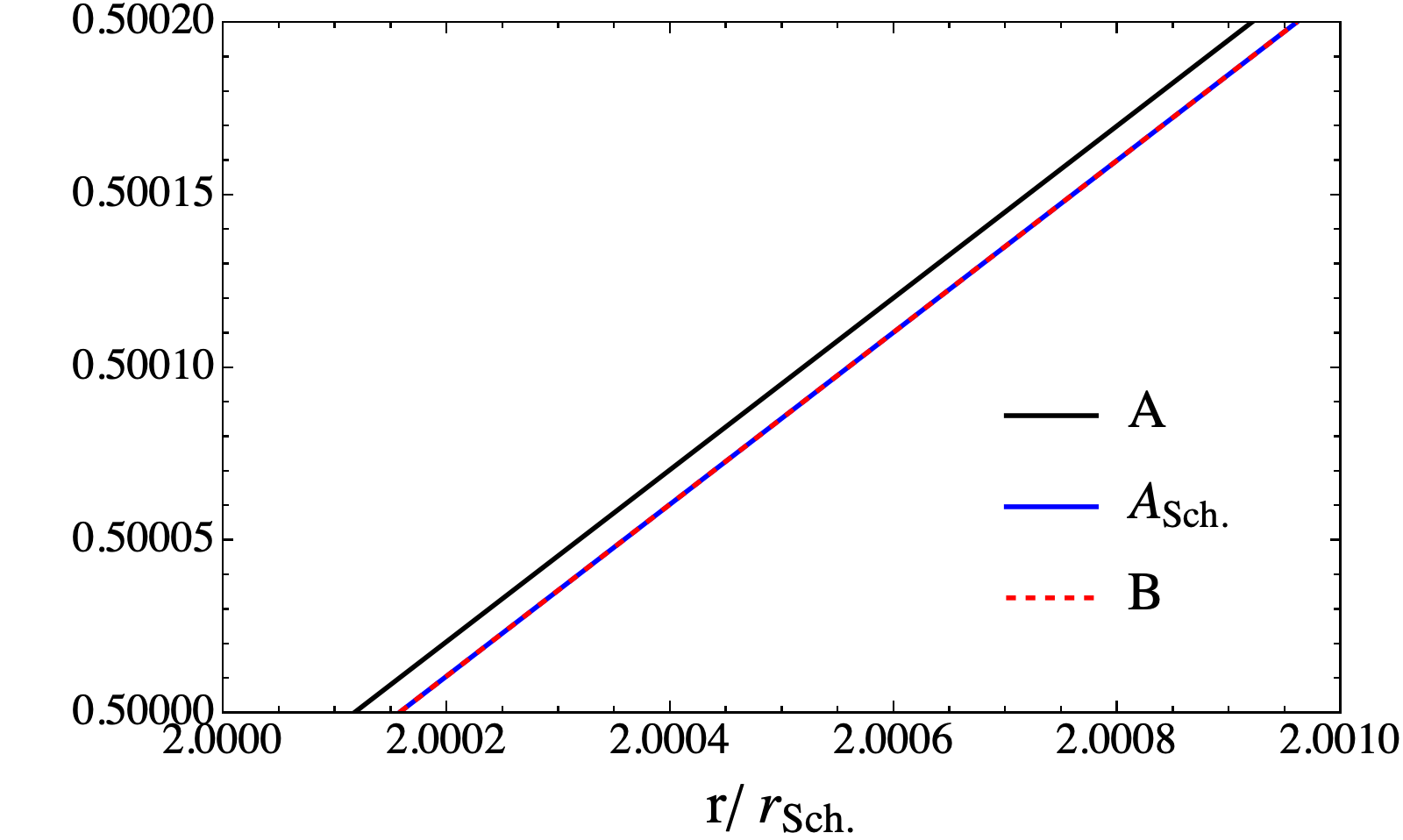}
    \begin{center}
    \includegraphics[scale=0.4]{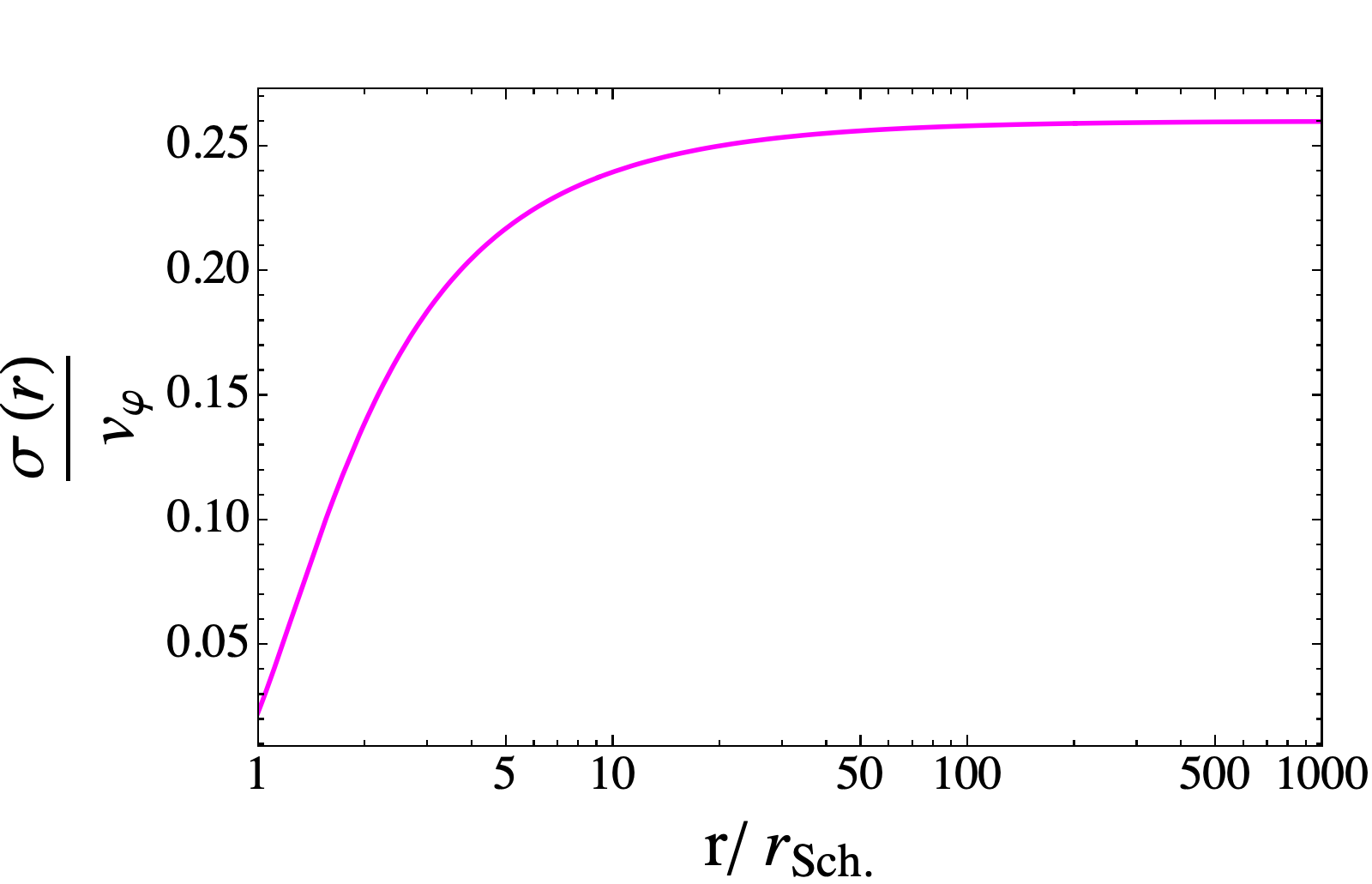}
    \end{center}
\caption{\textit{Left panel:} The metric functions $A(r)$ and $B(r)$ for the hairy black hole, as a function of the distance from the black hole horizon. \textit{Right panel:} Magnified plot of the metric functions, along with that for the Schwarzschild metric shown for comparision. \textit{Bottom panel:} The scalar field profile for the hairy black hole. In these plots, we consider $\lambda= 0.001$, ~$\alpha=1,~ \Lambda= 10^{-5}m_P$.} 
\label{fig:hbhsoln}  
\end{figure}

Now let us investigate the hairy black hole solutions. In the symmetry-broken phase of the black hole spacetime, the scalar field and the coupling function are better denoted as
\begin{align}
\varphi(r) = \bigg(v + \frac{\sigma(r)}{\sqrt{2}} \bigg)e^{i \theta(r)}, \qquad f(\sigma)=\frac{\alpha}{\Lambda_2^2}  \sigma ^2  + \frac{\sqrt{\alpha  \lambda }}{\Lambda_2\Lambda_1^2}\; \sigma ^3 +\frac{\lambda}{4\Lambda_1^4} \sigma ^4\, .
\end{align}
With new variables, the Einstein equations and scalar field equations are yielded as
\begin{align}
  &\frac{1}{\kappa^2} \bigg(R_{\mu \nu} - \frac{1}{2} R g_{\mu \nu} \bigg) =  \bigg[-\frac{1}{2} \nabla_{\gamma} \sigma \nabla^{\gamma} \sigma - \bigg(v + \frac{\sigma}{\sqrt{2}} \bigg)^2 \nabla_{\gamma} \theta \nabla^{\gamma} \theta \bigg]g_{\mu \nu} \nonumber\\
  &\qquad \qquad + \nabla_{\mu} \sigma \nabla_{\nu} \sigma + 2 \bigg(v + \frac{\sigma}{\sqrt{2}} \bigg)^2 \nabla_{\mu} \theta \, \nabla_{\nu} \theta - (g_{\rho \mu} g_{\lambda \nu} + g_{\lambda \mu} g_{\rho \nu}) \eta^{\kappa \lambda \alpha \beta} \tilde{R}^{\rho \gamma}{}_{\alpha \beta} \nabla_{\gamma} \nabla_{\kappa} f(\sigma), \\
  &\nabla^{\alpha} \nabla_{\alpha} \sigma - \sqrt{2} \bigg(v + \frac{\sigma}{\sqrt{2}}\bigg) \nabla^{\alpha} \theta\, \nabla_{\alpha} \theta + f_{\sigma}\, \mathcal{G} = 0, \qquad \bigg(v + \frac{\sigma}{\sqrt{2}} \bigg)  \nabla^{\alpha} \nabla_{\alpha} \theta + \sqrt{2}\, \nabla^{\alpha} \sigma \nabla_{\alpha} \theta = 0\,,
\end{align}
where $\tilde{R}^{\rho \gamma}{}_{\alpha \beta} = \eta^{\rho \gamma \sigma \tau} R_{\sigma \tau \alpha \beta} = \frac{\epsilon^{\rho \gamma \sigma \tau}}{\sqrt{-g}} R_{\sigma \tau \alpha \beta}$, 
and these are explicitly written as
\begin{align}
  &\frac{B'}{r B}-\frac{1}{r^2 B}+\frac{1}{r^2} +\frac{\kappa^2}{2}\left(\sigma '\right)^2 + \kappa^2 \left(\theta '\right)^2 \bigg(v+\frac{\sigma}{\sqrt{2}} \bigg)^2 \nonumber\\
  &\qquad \qquad -\frac{4 \kappa^2}{r^2} \left[\frac{(3 B-1) B' f_{\sigma} \sigma '}{B}+2 (B-1) \left(f_{\sigma \sigma} \left(\sigma '\right)^2+f_{\sigma} \sigma ''\right)\right] = 0, \\
  &\frac{A'}{r A}-\frac{1}{r^2 B}+\frac{1}{r^2} -\frac{\kappa^2}{2} \left(\sigma '\right)^2 - \kappa^2 \left(\theta '\right)^2 \bigg(v+\frac{\sigma}{\sqrt{2}}\bigg)^2 +\frac{4 \kappa^2 (1-3 B) A' f_{\sigma} \sigma '}{A r^2} = 0,\\
  &\frac{A''}{2 A}+\frac{B'}{4 B} \left(\frac{A'}{A}+\frac{2}{r}\right)+\frac{A' }{4 A} \left(\frac{2}{r}-\frac{A'}{A}\right)+ \frac{\kappa^2}{2} \left(\sigma '\right)^2 +\kappa^2 \left(\theta '\right)^2 \bigg(v+\frac{\sigma}{\sqrt{2}} \bigg)^2 \nonumber\\
  &\qquad \qquad + \frac{2 \kappa^2}{A r}  \left[\frac{A' }{A} \left(B A'-3 A B'\right)f_{\sigma} \sigma '-2 B \left(A' f_{\sigma \sigma} \left(\sigma '\right)^2+f_{\sigma} \left(A'' \sigma '+A' \sigma ''\right)\right)\right]= 0\,.
\end{align}
Here the field $\theta(r)$ is decoupled from the system and should be trivial in order to maintain the flux conservation, which is discussed in \cite{Latosh:2023cxm}. Thus we set $\theta = constant$. Since we seek for regular black hole solutions, we impose the boundary condition near the horizon as follows
\begin{align}
&A(r) \sim A_h \epsilon + \mathcal{O}(\epsilon^2) , \; \; \; B(r) \sim B_h \epsilon  + \mathcal{O}(\epsilon^2), \; \; \; \sigma(r) \sim \sigma_{h} + {\sigma_{h,1}} \epsilon  + \mathcal{O}(\epsilon^2) \,,\label{eq:NHexpSB}  
\end{align}
where the near horizon expansion of the metric and the scalar field $\sigma(r)$ are following 
\begin{align}
&B_h = \frac{r_h^3}{48 \kappa^2  (f_{\sigma _h})^2} \left(1-\sqrt{1-\frac{96 \kappa^2}{r_h^4} (f_{\sigma _h})^2}\right), \\
&\sigma_{h,1} =- \frac{r_h}{4 \kappa^2 f_{\sigma _h}} \left(1 - \sqrt{1- \frac{96 \kappa^2}{r_h^4} (f_{\sigma _h})^2 } \right) .
\end{align}
The boundary conditions are described only by the two independent variables $A_h$ and $\varphi_h$, along with the couplings $\alpha/\Lambda_2^2$ and $\lambda/\Lambda_1^4$. The regularity condition requires
\begin{align}
(f_{\sigma _h})^2 < \frac{r_h^4}{96 \kappa^2}. \label{eq:SMrgcnd}
\end{align}
We demonstrate our numerical result in Fig. \ref{fig:hbhsoln}, which is found by solving the above set of equations along with the boundary conditions, and using input parameters that are cosmologically consistent. The upper panel shows the metric functions $A(r)$ and $B(r)$ for hairy black holes as a function of $r/r_{\rm Sch.}$. The right panel shows a magnified plot for the same, where a small change can be seen, particularly for $A(r)$, compared to that of the Schwarzschild metric, which is shown alongside for comparison.  Note that here, we consider both black holes of the same size, i.e. $r_h=r_{\rm Sch.}$, which is chosen to be that of a black hole with $M_{\rm Sch.}=10~\text{g}$. The mass of the hairy black hole is found by evaluating the asymptotic limit of  $\frac{r^2 A'(r)}{2 G}$, and is found to be $10.0002 ~\text{g}$, deviating slightly from the Schwarzschild case. The lower panel shows the non-trivial profile for the scalar field,  which reaches an asymptotic value at large $r$. This scalar field profile is  found to be stable against the scalar field perturbations, which was verified in Ref. \cite{Hyun:2024sfv}, by analysing the quasinormal modes for the hairy black holes.

\section{Greybody factor}\label{Sec6}
%Since the PBHs that could develop scalar hair are in the ultralight mass range, they would evaporate away in the very early Universe much before BBN. Hence, they are not motivating as a dark matter candidate. However, the evanescence of these ultralight PBHs via Hawking radiation can play a role in producing dark matter \cite{Gondolo:2020uqv, Bernal:2020bjf}, baryon asymmetry \cite{} and relic gravitational waves \cite{}. Thus, it is important to understand the Hawking radiation of these non-Schwarzschild hairy PBHs. As a first step, here we calculate the Greybody factors of these hairy BHs numerically and investigate whether there is any difference from the Schwarzschild case. 

Since PBHs capable of developing scalar hair lie in the ultralight-mass range, they would evaporate through Hawking radiation in the very early Universe, well before BBN, and are therefore not viable dark matter candidates. Nevertheless, their evaporation may have important cosmological consequences, including the production of dark matter \cite{Fujita:2014hha, Lennon:2017tqq, Allahverdi:2017sks, Morrison:2018xla, Gondolo:2020uqv, Bernal:2020bjf,   Hooper:2019gtx, Chaudhuri:2020wjo, Masina:2020xhk, Baldes:2020nuv, Bernal:2020ili, Bernal:2020kse, Masina:2021zpu, Bernal:2021bbv, Bernal:2021yyb, Samanta:2021mdm, Sandick:2021gew, Cheek:2021odj, Cheek:2021cfe, Barman:2021ost, Borah:2022iym, Barman:2022gjo, Chen:2023lnj,Chen:2023tzd,Kim:2023ixo,Gehrman:2023qjn}, the generation of the baryon asymmetry \cite{Hawking:1974rv, Carr:1976zz,Baumann:2007yr, Hook:2014mla, Fujita:2014hha, Hamada:2016jnq, Morrison:2018xla, Hooper:2020otu, Perez-Gonzalez:2020vnz, Datta:2020bht, JyotiDas:2021shi, Smyth:2021lkn, Barman:2021ost, Bernal:2022pue, Barman:2022gjo, Ambrosone:2021lsx,Calabrese:2023key,Calabrese:2023bxz,Gehrman:2022imk,Gehrman:2023esa, Barman:2024slw}, and the production of a relic gravitational-wave background \cite{Anantua:2008am, Papanikolaou:2020qtd, Domenech:2020ssp, Domenech:2021wkk, Zagorac:2019ekv, Hooper:2020evu}. Although a comprehensive analysis of Hawking radiation is beyond the scope of this work and is left for future study, we investigate the associated greybody factors and compare the results for hairy and Schwarzschild black holes.  A recent study of greybody factors for scalar-tensor theories, can be found in Ref. \cite{Antoniou:2025bvg}.  

Greybody factors characterise the scattering process around a black hole background, and indicate the deviation of the Hawking spectrum from a perfect black body. To compute them, we consider perturbations of the scalar field $\varphi$ near the black hole spacetime. Its equation of motion and the corresponding mode ansatz are given by
\begin{align}
    &\Box \delta \varphi=0, \label{eq:greybdeom}\\
    &\delta \varphi (t,r, \theta, \Phi) = \int d \omega \frac{\psi(r)}{r} Y_l^{m} (\theta,\Phi) e^{-i \omega t}\,,
\end{align}
where $Y_l^{m} (\theta,\Phi)$ are the spherical harmonics and $\omega$ corresponds to the frequency of the partial waves. Considering the spherically symmetric black hole metric given in Eq. \eqref{eq:BHmetric}, the equation of motion (\ref{eq:greybdeom}) can be recast in the form of a Schrödinger-type equation, given as
\begin{align}
    \frac{d^2 \psi(r_*)}{d r_*^2} + [\omega^2 -V_{\rm eff} (r)]\psi(r_*) =0\,,\label{eq:schEOM}
\end{align}
using the tortoise coordinate $r_*$ defined as $\frac{dr_*}{d r}= (A B)^{-1/2}$.  The effective potential $V_{\rm eff} (r)$ is given by Eq. \eqref{eq:Veffr}. The asymptotic solutions to Eq. \eqref{eq:schEOM} are given by
\begin{align}
    &\psi (r_* \rightarrow-\infty) =  A_{\rm H} (\omega) e^{-i\omega r_*} + B_{\rm H} (\omega) e^{i\omega r_*}\,, \\ \nonumber
    &\psi (r_* \rightarrow\infty) = A_{\rm in} (\omega)e^{-i\omega r_*} + A_{\rm out}(\omega)e^{i\omega r_*}\,.
\end{align}
The greybody factors correspond to the transmission coefficients of a scattering process of an incoming test field ($\psi$ in this case), with a potential barrier due to the BH spacetime. We therefore impose a purely ingoing boundary conditions at the horizon, setting  $B_H(\omega) =0$, and normalize the wave function to unity, $A_H(\omega) =1$. The transmission coefficient or the greybody factor is then given by
\begin{align}
    \Gamma_l (\omega) = 1 -\frac{|A_{\rm out}(w)|^2}{|A_{\rm in}(w)|^2} = \frac{1}{|A_{\rm in}(\omega)|^2}\,,\label{eq:GF}
\end{align}
where in the second equality we used $|A_{\rm in}(w)|^2-|A_{\rm out}(w)|^2=1$, which follows from the flux conservation.

\begin{figure}[h]
  \includegraphics[scale=0.42]{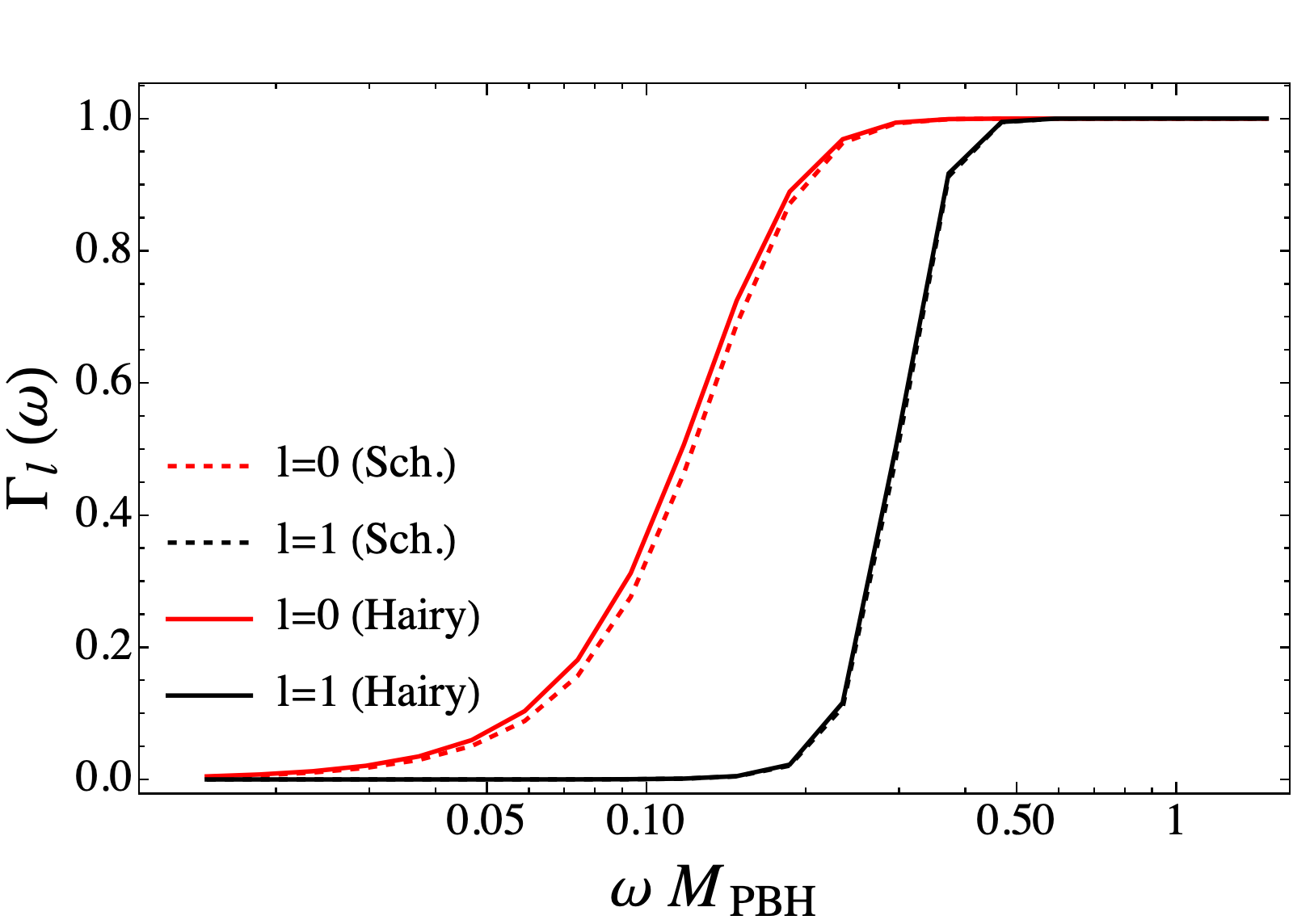}  ~~  \includegraphics[scale=0.42]{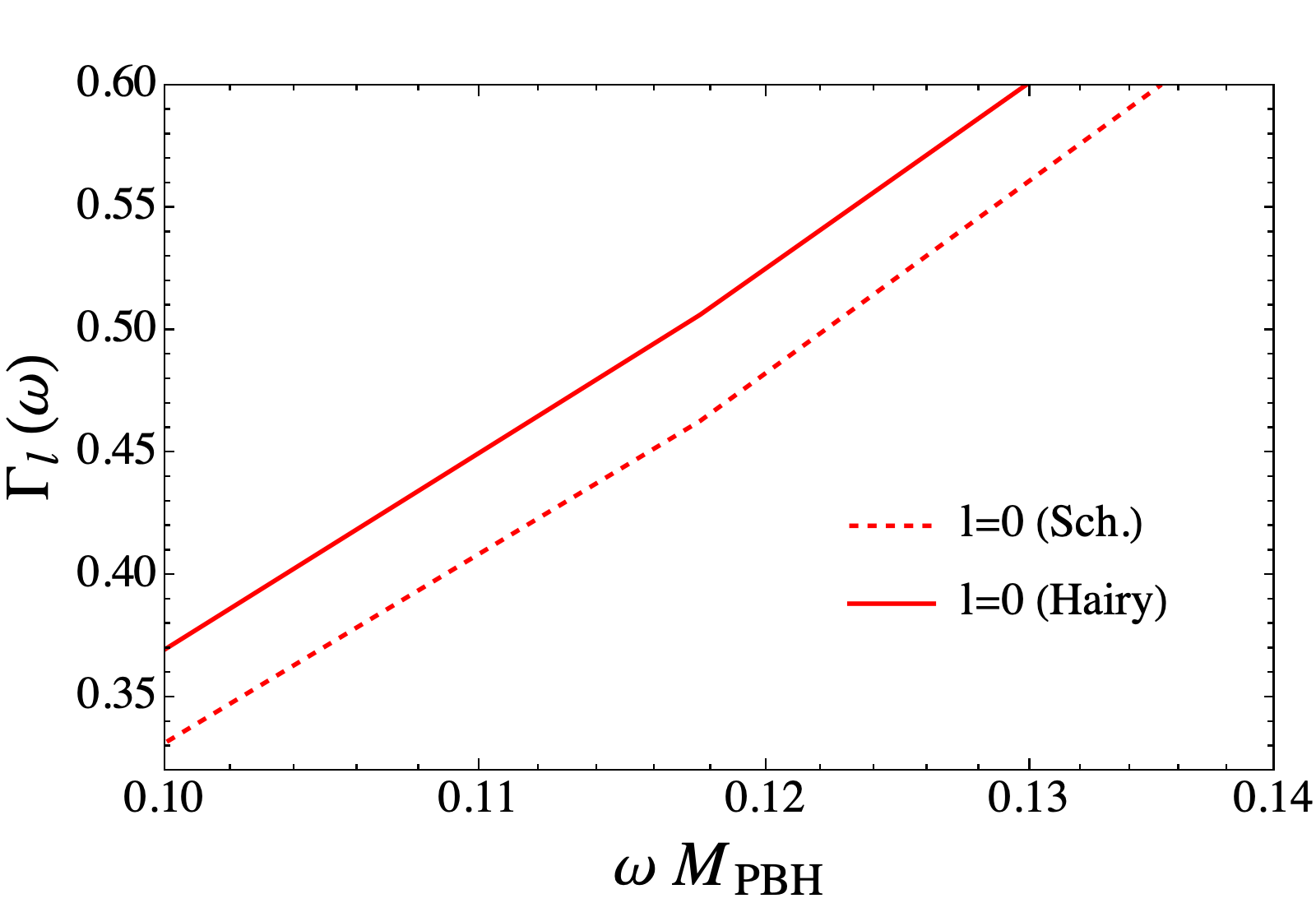}
\caption{\textit{Left panel:} The greybody factors for $l=0$ (red) and $l=1$ (black), for the Schwarzschild black hole (solid line) and hairy black hole (dashed line). \textit{Right panel:} Magnified plot for $l=0$. Here, we consider $M_{\rm Sch.} = 10$ g, $\alpha=1$, $\lambda= 10^{-3}$ and $\Lambda =10^{-5}~ m_P$.} 
\label{fig:GF}  
\end{figure}

By solving the equation for $\psi$ in both the Schwarzschild and the hairy black hole background subject to the above boundary conditions, we can
%We input the numerical values of the metric functions $A(r), B(r)$, along with the scalar hair $\sigma (r)$ for $V_{\rm eff}$  in Eq. \eqref{eq:schEOM}. The boundary conditions at the horizon discussed above are used as initial values to solve the equation. The asympotic solution at large $r$ is evaluated to 
extract the asymptotic coefficients $A_{\rm in} (\omega), A_{\rm out} (\omega)$ and find the corresponding greybody factor given by Eq. \eqref{eq:GF}. Fig. \ref{fig:GF} shows the greybody factors for $l=0,1$, as a function of the dimensionless quantity $ \omega M_{\rm PBH}$, considering the benchmark values: $M_{\rm Sch.} = 10$ g, $\alpha=1$, $\lambda= 10^{-3}$ and $\Lambda =10^{-5}~ m_P$, which satisfy the conditions discussed earlier for avoiding significant tachyonic growth. The solid lines indicate the greybody factors for the Schwarzschild case, while the dashed lines correspond to the hairy black hole. The right panel shows the magnified plot for $l=0$, where it is evident that the change in the greybody factor for the hairy black hole case is less than $\mathcal{O}(1)$. We observe a small deviation between the greybody factors of the hairy and Schwarzschild black holes, which becomes more pronounced for lower multipoles $l$. This behaviour can be understood from the structure of the effective potential, which is modified by the presence of scalar hair through its effect on the black hole geometry. For low multipoles, the centrifugal contribution is weak, being absent for $l=0$; and the transmission probability is therefore more sensitive to the hair-induced modifications of the near-horizon geometry. By contrast, at larger $l$, the effective potential is increasingly dominated by the centrifugal barrier, suppressing the relative effect of the scalar hair and causing the greybody factors to approach their Schwarzschild counterparts. 

Although the greybody factors of the massless scalar test field exhibit only small deviations from their Schwarzschild counterparts, this result alone may not imply that the full Hawking radiation spectra of the two black holes are similar. In particular, the substantial difference between their quasinormal mode spectra \cite{Hyun:2024sfv} indicates that the perturbative dynamics can be highly sensitive to the presence of scalar hair.   Our result therefore indicates only that the transmission probabilities for the particular test field and parameter range considered here are weakly affected by the scalar hair. A comprehensive comparison of the Hawking radiation spectra of these black holes would therefore be an interesting direction for future work, which is currently under investigation. Especially, going beyond the test-field approximation to account for backreaction is important for a more complete understanding of Hawking evaporation in these hairy black holes.

\section{Summary and Conclusion}\label{Sec7}

We investigated whether hairy black holes generated through spontaneous symmetry breaking in Einstein–Scalar–Gauss–Bonnet (ESGB) theory, with a complex scalar field respecting a global U(1) symmetry, can be compatible with cosmological evolution. In such a theory, Schwarzschild black holes can be unstable, and so undergo a phase transition to stable hairy black holes in the symmetry-broken phase via spontaneous symmetry breaking. This construction is particularly interesting because it provides the first realization of such a mechanism in an asymptotically flat spacetime. To establish closer connections with phenomenology and observations, it is therefore crucial to determine whether these hairy black holes can persist throughout cosmological history and, if so, to reconstruct their formation history and characterize any present-day remnants or observable signatures.

To this end, we extended the ESGB theory by introducing a scalar self-interaction that becomes relevant on cosmological scales but is assumed to be negligible near the black hole. We further assumed that the spacetime is well approximated by a black hole geometry out to approximately $10^3$ times the black hole radius, while approaching an FLRW geometry on larger scales. In contrast to the nearly static black hole background, extending the ESGB effective potential to the dynamical FLRW background has nontrivial consequences for the evolution of the scalar field throughout cosmological history. In particular, for scalar–GB couplings compatible with hairy black hole formation, the effective potential remains in the spontaneously broken phase during inflation. However, after inflation, the GB term changes sign in the FLRW background, temporarily rendering the effective potential unbounded from below. As the GB contribution decreases during cosmological evolution, the scalar self-interaction eventually becomes dominant and restores the  symmetry.

Within this schematic framework, we derived detailed constraints on the coupling strengths and the black hole mass. Firstly, requiring the scalar field to remain localized around its vacuum expectation value during inflation imposes stringent constraints on both the cutoff scale associated with the GB interactions and the black hole mass, as illustrated in Fig. \ref{fig:bndplt}. We then analysed the scalar-field perturbations after inflation. During the post-inflationary phase in which the effective potential is unbounded from below, these perturbations can undergo tachyonic amplification, leading to a substantial enhancement of the scalar field energy density and potentially disrupting the standard cosmological evolution. Avoiding efficient tachyonic growth further narrows the allowed parameter space and requires the cutoff scale to lie close to \(\Lambda\sim10^{-5}\,m_{\rm P}\). For cutoff scales around this range, only ultralight black holes with masses in the ballpark of around $1~\text{g}$-$10~\text{g}$ can develop scalar hair, motivating our investigation of such hairy primordial black holes. 

Since our analysis concerns the early Universe, the characteristic scale of a PBH can be comparable to the cosmological expansion scale, particularly around the time of its formation. As the Universe subsequently expands, the physical size of the newly formed black hole remains approximately fixed over the timescales relevant to our analysis, whereas the Hubble radius continues to grow. We therefore estimate the spatial and temporal regimes in which the spacetime near the black hole can be reliably approximated by the Schwarzschild metric, thereby ensuring the validity of our analysis. As shown in Fig. \ref{fig:zsch}, we find that the Schwarzschild approximation remains valid upto \(r\sim10^{2\text{--}3}r_{\rm Sch.}\), with this domain increasing as the Universe expands.\footnote{By contrast, for a stellar-mass astrophysical black hole, the horizon radius is of order kilometres, whereas the present-day cosmological expansion scale is of order gigaparsecs, establishing a clear separation between the two scales.}

%However, these two length scales are comparable in the early Universe for PBHs, especially during its formation time. We estimate the length and time scales, at which we can adequately describe the spacetime by the Schwarzschild metric close to the black hole, such that our analysis holds. This is presented in Fig. \ref{fig:zsch}, where we find that our estimation can safely hold until $r\sim 10^{2-3} r_{\rm Sch.}$, which keeps increasing as the  Universe expands. 

Now, at some point during this evolution, the symmetry of the vacuum is restored in the FLRW spacetime, after which, the scalar field with inefficient tachyonic growth, can undergo oscillation around $\phi=0$. The Schwarzschild configuration, with $\phi (r=r_{\rm Sch.})=0$, can however become unstable,  and rapidly undergo a phase transition to a hairy black hole characterized by $\phi(r=r_{\rm Sch.})=v_\phi$. We find that this transition occurs almost instantaneously on cosmological timescales, with $z_h\sim z_0$.  We obtain the resulting hairy black hole solutions numerically by solving the coupled Einstein and scalar-field equations, and present them in Fig. \ref{fig:hbhsoln}. Finally, we calculated the greybody  factor for the hairy black hole, where we find a small deviation from the Schwarzschild case, less than $\mathcal{O}(1)$. Although this change is small, a complete detailed analysis of the Hawking radiation spectrum is essential to determine the fate of the scalar hair of the black hole, and look for any unique imprints of the scalar hair in observations. We plan to investigate this in a subsequent study. This can essentially lead to plenty of rich phenomenology, such as production of relics, including dark matter, the baryon asymmetry, gravitational waves etc. via Hawking radiation, with possible signatures of the scalar hair. Additionally, it is also worth mentioning that a memory-burden effect \cite{Dvali:2018xpy, Dvali:2018ytn, Dvali:2024hsb, Alexandre:2024nuo, Thoss:2024hsr} can lead to a suppression to this evaporation, extending the lifetime of the ultralight PBHs to a late time, depending on the entropy suppression factor or model parameters. This might open interesting observational windows for the hairy ultralight PBHs.

%At some point during this evolution, the symmetry \cmb{of the vacuum} is restored in the FLRW spacetime, and the scalar field, with inefficient tachyonic growth, can undergo oscillation around $\phi=0$. However, close the Schwarzschild black hole horizon, scalar field with $\phi (r=r_{\rm Sch.})=0$ would be unstable and undergo a phase transition, almost instantly with $z_h \sim z_0$, to hairy black holes with $\phi (r=r_{\rm Sch.})=v_{\phi}$. The hairy black solutions are found numerically by solving the Einstein and scalar field equations in the black hole spacetime, and are presented in Fig. \ref{fig:hbhsoln}. Finally, we calculated the Greybody  factors for the hairy black hole, where we find a small deviation from the Schwarzschild case, less than $\mathcal{O}(1)$.

%While the key takeaway of our work is to point out that a cosmologically consistent modified gravity theory like ESGB, might open the possibility of PBHs with extra features in terms of scalar hair; this requires a further detailed and careful investigation. 

 To conclude, the central implication of our work is that a cosmologically viable modified-gravity framework, such as ESGB theory, may allow PBHs to possess scalar hair. Establishing this possibility conclusively, however, requires further careful investigation. Importantly, in such a scenario, the scalar profile at distances far away from the black hole ($r\gtrsim10^{2-3}r_{\rm Sch.}$) should be matched properly with the cosmological solution. With the scalar field oscillating around $\phi=0$ in the FLRW background with amplitude around $v_{\phi}a(z_0)/a(z)$, the profile is expected to be dynamical around the boundary $r\sim10^{2-3}r_{\rm Sch.}$, and requires a detailed numerical simulation or a better approximation of the metric of black holes in an expanding Universe. Additionally, a rigorous understanding of the fully non-linear dynamics during the tachyonic phase is important for providing more robust constraints, which will require dedicated lattice simulations.
%Additionally, it would be important to rigorously understand the full non-linear dynamics during the tachyonic phase, which requires a full lattice simulation. We also can consider, for instance, in the presence of a bare mass term large enough, the coherent oscillations at later epochs  might play the role of dark matter \cite{Laverda:2026slq}.  Such a mass term, if strong enough, can also modify the hairy black hole solutions \cite{}. Additionally, in the presence of an explicit symmetry-breaking term, there can be non-trivial dynamics along the angular $\theta$ direction of the complex field, that can play role in baryogenesis \cite{Liang:2019fkj}. We leave such interesting implications connecting these cosmological issues with black hole features, under the same modified gravity theory, for future studies.
 An interesting extension of our current scenario would be to introduce a sufficiently large bare mass term, in which case coherent scalar-field oscillations at later epochs could constitute a dark-matter component \cite{Laverda:2026slq}. Such a mass term, if strong enough, could also significantly modify the hairy black hole solutions \cite{Macedo:2019sem}. Furthermore, an explicit symmetry-breaking term could induce nontrivial dynamics along the angular or phase direction \(\theta\) of the complex scalar field, potentially providing a mechanism for baryogenesis \cite{Liang:2019fkj}. We leave a detailed investigation of these possibilities, and of the broader interplay between cosmological dynamics and black hole phenomenology within a unified modified-gravity framework for future work.

\vspace{0.2cm}
\section*{Acknowledgement}
 S.J.D. is supported by the National Science Centre
(Poland), under the research project no. \\ 2023/49/B/ST2/00856. S.J.D. was also supported by Institute for Basic Science under the project code, IBS-R018-D1.  M.P. was supported by the Institute for Basic Science (Grant No. IBS-R018-Y1) and by the National Research Foundation of Korea (Grant No.: RS-2026-25572374). We appreciate APCTP for its hospitality during completion of this work.
\bibliographystyle{JHEP}
\bibliography{ref} 

\end{document}